\documentclass[%
 reprint,
superscriptaddress,
 amsmath,amssymb,
 aps,
]{revtex4-1}

\usepackage{graphicx}% Include figure files

\usepackage{dcolumn}% Align table columns on decimal point
\usepackage{bm}% bold math
\usepackage{float}

\RequirePackage[T1]{fontenc}
\RequirePackage[utf8]{inputenc}
\RequirePackage{lmodern}
\newcommand {\ket} [1] {|{ #1 \rangle}}
\newcommand {\bra} [1] {\langle{ #1 |}}
\begin{document}

\title{Electron spin relaxation of single phosphorus donors in metal-oxide-semiconductor nanoscale devices}% Force line breaks with \\

\author{Stefanie B. Tenberg}
 \affiliation{Centre for Quantum Computation and Communication Technology, School of Electrical Engineering \& Telecommunications, UNSW Sydney, New South Wales 2052, Australia.}
 \author{Serwan Asaad}
 \affiliation{Centre for Quantum Computation and Communication Technology, School of Electrical Engineering \& Telecommunications, UNSW Sydney, New South Wales 2052, Australia.}
 \author{Mateusz T. M\k{a}dzik}
 \affiliation{Centre for Quantum Computation and Communication Technology, School of Electrical Engineering \& Telecommunications, UNSW Sydney, New South Wales 2052, Australia.}
 \author{Mark A. I. Johnson}
 \affiliation{Centre for Quantum Computation and Communication Technology, School of Electrical Engineering \& Telecommunications, UNSW Sydney, New South Wales 2052, Australia.}
 \author{Benjamin Joecker}
 \affiliation{Centre for Quantum Computation and Communication Technology, School of Electrical Engineering \& Telecommunications, UNSW Sydney, New South Wales 2052, Australia.}
\author{Arne Laucht}
 \affiliation{Centre for Quantum Computation and Communication Technology, School of Electrical Engineering \& Telecommunications, UNSW Sydney, New South Wales 2052, Australia.}
 \author{Fay E. Hudson}
 \affiliation{Centre for Quantum Computation and Communication Technology, School of Electrical Engineering \& Telecommunications, UNSW Sydney, New South Wales 2052, Australia.}
  \author{Kohei M. Itoh}
 \affiliation{School of Fundamental Science and Technology, Keio University, 3-14-1 Hiyoshi, 223-8522, Japan.}
   \author{A. Malwin Jakob}
  \affiliation{Centre for Quantum Computation and Communication Technology, School of Physics, University of Melbourne, Melbourne, Victoria 3010, Australia.}
     \author{Brett C. Johnson}
  \affiliation{Centre for Quantum Computation and Communication Technology, School of Physics, University of Melbourne, Melbourne, Victoria 3010, Australia.}
   \author{David N. Jamieson}
 \affiliation{Centre for Quantum Computation and Communication Technology, School of Physics, University of Melbourne, Melbourne, Victoria 3010, Australia.}
  \author{Jeffrey C. McCallum}
 \affiliation{Centre for Quantum Computation and Communication Technology, School of Physics, University of Melbourne, Melbourne, Victoria 3010, Australia.}
 \author{Andrew S. Dzurak}
 \affiliation{Centre for Quantum Computation and Communication Technology, School of Electrical Engineering \& Telecommunications, UNSW Sydney, New South Wales 2052, Australia.}
\author{Robert Joynt}
 \affiliation{University of Wisconsin-Madison, Physics Department
1150 University Ave, Madison, Wisconsin 53706, USA, and Kavli Institute for Theoretical Sciences, University of Chinese Academy of Sciences, Beijing 100190, China}
\author{Andrea Morello}
 \email{a.morello@unsw.edu.au}
 \affiliation{Centre for Quantum Computation and Communication Technology, School of Electrical Engineering \& Telecommunications, UNSW Sydney, New South Wales 2052, Australia.}

\date{\today}

\begin{abstract}
We analyze the electron spin relaxation rate $1/T_1$ of individual ion-implanted $^{31}$P donors, in a large set of metal-oxide-semiconductor (MOS) silicon nanoscale devices, with the aim of identifying spin relaxation mechanisms peculiar to the environment of the spins. The measurements are conducted at low temperatures ($T\approx 100$~mK), as a function of external magnetic field $B_0$ and donor electrochemical potential $\mu_{\rm D}$. We observe a magnetic field dependence of the form $1/T_1\propto B_0^5$ for $B_0\gtrsim 3\,$ T, corresponding to the phonon-induced relaxation typical of donors in the bulk. However, the relaxation rate varies by up to two orders of magnitude between different devices. We attribute these differences to variations in lattice strain at the location of the donor.
For $B_0\lesssim 3\,$T, the relaxation rate changes to $1/T_1\propto B_0$ for two devices. This is consistent with relaxation induced by evanescent-wave Johnson noise created by the metal structures fabricated above the donors. At such low fields, where $T_1>1\,$s, we also observe and quantify the spurious increase of $1/T_1$ when the electrochemical potential of the spin excited state $\ket{\uparrow}$ comes in proximity to empty states in the charge reservoir, leading to spin-dependent tunneling that resets the spin to $\ket{\downarrow}$. These results give precious insights into the microscopic phenomena that affect spin relaxation in MOS nanoscale devices, and provide strategies for engineering spin qubits with improved spin lifetimes.
%\begin{description}


%\end{description}
\end{abstract}

\maketitle

\section{\label{sec:introduction}Introduction}

Electrons bound to shallow donors in silicon became a centerpoint of solid-state physics in the 1950s, when the study of their spin and orbital states was used as a benchmark for the then emerging theories of band structure, effective mass and impurity states in solids \cite{Kohn1955}. In particular, the detailed analysis of the donor electron spin-lattice relaxation time $T_1$ provided key insights into the multi-valley band structure of silicon, and the way it influences spin-phonon coupling \cite{Wilson1961}.

Fast-forward half a century, donor spins have become the subject of intense research for their potential use in quantum computing \cite{Kane1998,Hill2015,Pica2016,Tosi2017}. In this context, the old results on the electron spin $T_1$ seemed to provide ample reassurance that spin lifetime would not constitute a limitation to the encoding and protection of quantum information. The donor electron $T_1$ in bulk samples exceeds an hour at cryogenic temperatures and moderate magnetic fields \cite{Feher1959}, whereas the spin decoherence time $T_2$ is limited to a few hundred microseconds  \cite{Gordon1958,Pla2012}, due to the coupling of the electron spin to the bath of spin-1/2 $^{29}$Si nuclei present with 4.7\% abundance in natural silicon. However, the adoption of isotopically enriched $^{28}$Si samples, where the concentration of $^{29}$Si nuclei is reduced below 0.1\% \cite{Itoh2014}, has allowed extending $T_2$ close to \cite{Muhonen2014} or beyond \cite{Tyryshkin2012} one second. This comes within an order of magnitude of the $T_1$ time observed in nanoscale single-donor qubit devices \cite{Morello2010} at the magnetic fields $\gtrsim 1$~T typically used for control and readout of the electron spin \cite{Pla2012}, and calls for an effort to understand in detail all spurious channels of spin relaxation. 

In this work, we provide an extensive collection of experimental results and theoretical models on the electron spin relaxation time $T_1$ of single $^{31}$P donors in silicon metal-oxide-semiconductor (MOS) nanoelectronics devices, with the aim of elucidating how the environment of the donors influences the spin lifetime. Earlier measurements of $T_1$ on single donors in nanoscale devices \cite{Morello2010,Tracy2013,Watson2015,Weber2018} had already shown evidence of deviation from bulk-like behavior. Here, by analyzing data on 7 different devices, we uncover several microscopic mechanisms that affect the spin relaxation time. In particular, we provide evidence for relaxation induced by evanescent-wave Johnson noise (EWJN), by electron tunneling to a nearby reservoir, and modifications of the spin-phonon relaxation rate caused by strain.

The paper is organized as follows. Sec. \ref{sec:background} gives an overview of the theory of electron spin relaxation of donors in silicon, covering both bulk effects (phonon-induced relaxation) and phenomena specific to donors near metallic nanostructures (evanescent-wave Johnson noise, charge noise). Sec. \ref{sec:qubit} describes the details of our physical system and the experimental setup, as well as the measurement protocols used to acquire the data. Sec. \ref{sec:extB} shows the magnetic field dependence of $T_1$ in several devices, both in natural and isotopically-enriched silicon, with a detailed analysis of the low-field (Sec. \ref{sec:ewjn}) and high-field (Sec. \ref{sec:phonon}) relaxation channels. Sec. \ref{sec:cotunnelling} presents evidence of spin relaxation caused by tunneling to a nearby charge reservoir. Finally, Sec. \ref{sec:conclusion} discusses the results and the remaining open questions. 

\section{Background} \label{sec:background}

We describe a single $^{31}$P donor in silicon, subjected to an external magnetic field $B_0 \parallel \hat{z}$, with the following spin Hamiltonian:
\begin{equation}
    \mathcal{H}_{\rm P}= g_z \mu_{\rm B}B_0 S_z - h \gamma_n B_0 I_z + hA \mathbf{S}\cdot \mathbf{I}, \label{eq:HP}
\end{equation}
where $h$ is the Planck constant, $g_z$ is the component the electron Land{\`e} g-tensor along the field direction, $\mu_{\rm B}$ is the Bohr magneton, $\gamma_n=17.25$~MHz/T is the nuclear gyromagnetic ratio, $A$ is the electron-nuclear hyperfine coupling, $\mathbf{S}$ and $\mathbf{I}$ are spin-1/2 vector Pauli matrices describing the electron and the $^{31}$P nuclear spins, respectively, and $S_z, I_z$ are the operators representing the electron and nuclear spin projections along the $\hat{z}$-axis. For $^{31}$P donors in bulk silicon, the parameters in Eq.~\ref{eq:HP} take the values $g_z=1.9985$ (corresponding to $g_z \mu_{\rm B}/h = 27.971$~GHz) and $A=117.53$~MHz, but the distortion of the wavefunction caused by electric fields, strain or local confinement can result in small shifts of such values \cite{Laucht2015}.

In this paper we focus on the physics of the electron spin alone. Earlier experiments on the $^{31}$P nucleus \cite{Pla2013} have shown that it retains its state for extremely long times (typically many days, or even months). Moreover, we work in the regime where the electron Zeeman energy $g\mu_{\rm B}B_0$ greatly exceeds the hyperfine coupling $A$, and the electron-nuclear eigenstates are simply the tensor products of the electron ($\ket{\downarrow},\ket{\uparrow}$) and nuclear ($\ket{\Downarrow},\ket{\Uparrow}$) basis states. Therefore, choosing for example to prepare the nuclear spin laways in the $\ket{\Uparrow}$ state, the donor Hamiltonian can be truncated to an electron-only operator:
\begin{equation}
\mathcal{H} = (g_z\mu_{\rm B}B_0 + hA/2) S_z, \label{eq:Helectron}
\end{equation}
where the term $hA/2$ has the only effect of adding a small contribution to the electron spin energy splitting. This is inconsequential for the discussion of electron spin relaxation, and will be ignored from here onward.

Electron spin relaxation consists of transitions between the $\ket{\uparrow}$ and $\ket{\downarrow}$ basis states leading to thermal equilibrium with a bath at temperature $T$, and is mathematically described by the presence of off-diagonal matrix elements in the Hamiltonian, coupling the spin to some operators of the bath. In a simplified picture, we can describe the bath as a noise source that introduces a perturbation to the Hamiltonian described by: 
\begin{equation} \label{eq:pertH}
\mathcal{H'}[\lambda(t)]=\bm{\Delta}_{\perp}[\lambda(t)] \bm{S}.
\end{equation}
Here $\bm{\Delta}_{\perp}[\lambda(t)]$ is an operator that does not commute with $\mathcal{H}$, and depends on the parameter $\lambda(t)$ which describes the noise acting on the electron spin.
The electron relaxation rate is the sum of the decay ($W_{\uparrow\downarrow}$) and excitation ($W_{\downarrow\uparrow}$) rates: 
\begin{equation} \label{eq:emission}
T_1^{-1}(\lambda)=W_{\uparrow\downarrow}+W_{\downarrow\uparrow}.
\end{equation}
Thermal equilibrium is obtained by imposing that decay and excitation rates obey the detailed balance condition:
\begin{equation}
    \frac{W_{\downarrow\uparrow}}{W_{\uparrow\downarrow}}=\exp\left(-\frac{g_z \mu_{\rm B}B_0}{k_{\rm B}T}\right).
\end{equation}
In the experiments presented here, conducted at $B_0 > 0.5$~T and $T\approx 200$~mK, $g_z \mu_{\rm B}B_0\gg k_{\rm B}T$ and we can approximate $T_1^{-1}(\lambda)\approx W_{\uparrow\downarrow}$, with:
\begin{equation} \label{eq:fgr}
W_{\uparrow\downarrow}=\frac{2\pi}{\hbar}\left|\bra{\downarrow}\mathcal{H'}[\lambda(t)]\ket{\uparrow}\right|^2\rho_{\rm f}.
\end{equation}
This expression is an application of Fermi's golden rule, where $\rho_{\rm f}$ is the density of available final states for emission of energy from the spin into the bath. 
Introducing the transition operator of the noise perturbation 
\begin{equation} \label{eq:transitionmoment}
\bm{D}_{\perp,\lambda}=\frac{\partial \mathcal{H'}[\lambda(t)]}{\partial \lambda}
\end{equation}
 and the noise power spectral density
\begin{equation} \label{eq:density}
S_\lambda\left(\omega\right) = \int_{-\infty}^{+\infty}d\tau \langle \lambda(0)\lambda(\tau)\rangle \exp(-i\omega\tau),
\end{equation}
we can express the total relaxation rate as \cite{Yan2016, Cottet2003}  
\begin{equation} \label{eq:t1general}
T_1^{-1}=\sum_\lambda \frac{|\bra{\uparrow} \bm{D}_{\lambda, \perp} \ket{\downarrow}|^2}{\hbar^2} S_\lambda\left(\omega_0\right).
\end{equation}

\subsection{Phonon-induced relaxation}

In bulk silicon, the dominant mechanism that creates a transverse operator $\bm{\Delta}_\perp[\lambda(t)]$ acting on the donor electron spin is the modification of the $g$-tensor caused by elastic distortions of the crystal lattice (phonons). 

The band structure of silicon contains six degenerate conduction band minima along directions $\pm x, \pm y, \pm z$ (labeled below by the index $j=1,2, \ldots ,6$) at finite crystal momentum $k_0$, called valleys. A bound electron state in silicon must be constructed from linear combinations of the 6 valleys, whose index effectively constitutes an additional quantum number, in addition to the usual hydrogen-like principal, orbital and magnetic quantum numbers. The spherical symmetry of the Coulomb potential produced by the donor nucleus is broken by the cubic crystal field potential, creating a valley-orbit coupling. As a result, the ground $1s$ orbital state is further split into six valley-orbit states: a singlet with $A_1$ symmetry (ground state), a triplet with $T_2$ symmetry and a doublet with $E$ symmetry, with wave functions $\Psi_i=\sum_{j=1}^6\alpha_i^{(j)}\psi^{(j)}$, where $\psi^{(j)}$ are envelope-modulated Bloch functions of the $1s$ orbital and \cite{Kohn1955, Saraiva2015}
\begin{subequations} 
\begin{align} 
\alpha_1^{(j)}=\frac{1}{\sqrt{6}}(1, 1, 1, 1, 1, 1),\ \ (A_1)\\
\alpha_2^{(j)}=\frac{1}{\sqrt{2}}(1, -1, 0,0,0,0),\ \ (T_2)\\
\alpha_3^{(j)}=\frac{1}{\sqrt{2}}(0,0,1,-1,0,0),\ \ (T_2)\\
\alpha_4^{(j)}=\frac{1}{\sqrt{2}}(0,0,0,0,1,-1).\ \ (T_2)\\
\alpha_5^{(j)}=\frac{1}{2}(1, 1, -1, -1, 0, 0),\ \ (E)\\
\alpha_6^{(j)}=\frac{1}{2}(1, 1, 0, 0, -1, -1).\ \ (E)
\end{align} \label{eq:valleys}
\end{subequations}

When a phonon with wave vector $\mathbf{q}$ travels through the crystal, it creates a local strain $\vec{U}$ that inhomogeneously deforms the lattice by the displacement 
\begin{equation}
\mathbf{Q}(\mathbf{r})=\sum_{q,t}\left[\mathbf{e}_t(\mathbf{q})a_{\mathbf{q},t}e^{i\mathbf{q}\cdot\mathbf{r}}+\mathbf{e}_t^*(\mathbf{q})a_{\mathbf{q},t}^* e^{-i\mathbf{q}\cdot\mathbf{r}}\right],
\end{equation}
where $\mathbf{e}(\mathbf{q})=\mathbf{e}^*(\mathbf{-q})$ is the polarization vector, $a_{\mathbf{q},t}$ the displacement amplitude and $t=x,y,z$. The deformation alters the crystal symmetry such that the $j$th valley is shifted by an energy
\begin{equation} \label{eq:energyshift}
\epsilon^{(j)}=\sum_{t,t'}U_{t,t'}\left(\Xi_d\delta_{t,t'}+\Xi_u G_t^{(j)}G_{t'}^{(j)}\right),
\end{equation}
where $U_{t, t'}$ is the component of the strain tensor $\vec{U}$, $\mathbf{G}^{(j)}$ is the unit vector pointing from the origin to the bottom of the $j$th valley in the first Brillouin zone, and $\Xi_d$ and $\Xi_u$ are the Herring deformation-potential which describe the shift in the band edge energy caused by isotropic dilations and uniaxial strain, respectively \cite{Herring1956, Hasegawa1960}. 
If unperturbed, the ground state $A_1$ (Eq. \ref{eq:valleys}) has an equal population of all valleys. As a consequence of the energy shifts $\epsilon^{(j)}$ caused by the lattice phonon, the relative valley populations become unequal, causing the mixing of some excited states with the ground state. This effect is called "valley-repopulation" and causes a change in the electron $g$-factor. 

The $g$-factor of each valley depends on the spin-orbit interaction, which differs whether the electron moves in or out of plane with respect to the external magnetic field, resulting in an anisotropic value given by \cite{Wilson1961}:
\begin{equation}
g^2=g_{||}^2\cos^2\theta+g_\perp^2\sin^2\theta,
\end{equation}
where $\theta$ is the angle between $B_0$ and the valley axis and $g_{||}$ and $g_\perp$ are the $g$ values with $B_0$ pointing parallel and perpendicular to the
valley axes, respectively. In the unperturbed case, once averaged over all valley states according to their population, the g-factor actually becomes isotropic for the $A_1$ ground state due to the even valley population:
\begin{equation}
g_0=\frac{1}{3}g_{||}+\frac{2}{3}g_\perp.
\end{equation}

 However, in the strained case, the valley population is unequal which leads to an anisotropic $g$ which depends on the amount of strain. For instance, for stress along the $[100]$ direction, the $g$-factor becomes \cite{Wilson1961}:
 \begin{multline}
 g-g_0=\frac{1}{6}\left(g_{||}-g_\perp\right)\left(1-\frac{3}{2}\sin^2\theta\right)\\
 \times\left[1-\left(1+3x/2\right)\sqrt{1+x/3+x^2/4}\right],
\end{multline}
with $x=\Xi_u'/E_{12}$, where $\Xi_u'$ is the deformation potential adjusted for stress and $E_{12}$ is the valley-orbit splitting between the ground state $A_1$ and the doublet state $E$. 
This $g$-factor anisotropy effectively couples the electron spin $\mathbf{S}$ to the lattice phonon $\mathbf{q}$ via the Hamiltonian  \cite{Hasegawa1960}:
\begin{equation} \label{eq:ephH}
\mathcal{H'}_{\rm ph} = \frac{2g'\mu_B B_0 \Xi_u}{-3E_{12}}f(q)q\left(a_{\mathbf{q},t}\sum_{r}D_r^{(t)}D_r^{(t')}+c.c.\right)S_{t'} 
\end{equation}
where $g'=\frac{1}{3}(g_{||}-g_\perp)$, $f(q)=1/\left[1+\frac{1}{4}a_0^{*2}q^2\right]^2$, $a_0^{*}$ is the effective Bohr radius, and 
\begin{equation} \label{eq:couplMatrix}
D_r=3\sum_j\alpha^{(j)}\alpha_r^{(j)} \mathbf{U}^{(j)}
\end{equation}
is a tensor that describes the geometrical structure of the conduction band edge, with $r$ labeling the valley-orbit excited states and $\mathbf{U}^{(j)}$ the tensor that selects the direction of the $j$-th valley.

The spin-phonon interaction described by Eq.~\eqref{eq:ephH} represents one example of off-diagonal perturbation $\mathcal{H}'$ as in the general formalism of Eq.~\eqref{eq:fgr}. From this, Hasegawa \cite{Hasegawa1960} calculated the donor spin-lattice relaxation rate as:
 \begin{equation} \label{eq:fullT1}
\begin{aligned} 
T_{1,\rm rp}^{-1}  & =   \frac{1}{90\pi}\left(\frac{g_{||}-g_\perp}{g_0}\right)\left(\frac{\Xi_u}{E_{12}}\right)^2\left(\frac{1}{\rho v_t^5}+\frac{2}{3\rho v_l^5}\right)\\
&  \times \left(\frac{g\mu_{\rm B}B_0}{\hbar}\right)^4f_{\rm rp}(\theta)\cdot k_{\rm B} T \\
& =  K_4^{\rm rp}B_0^4T,
\end{aligned}
 \end{equation}
where $v_t = 5860$~m/s and $v_l = 8480$~m/s are the transverse and longitudinal sound velocities in silicon, respectively, $\rho=2330$~kg/m$^3$ is the density of silicon and $f_{\rm rp}(\theta)=\sin^2\theta(1+3\cos^2\theta)$ is a geometric factor where $\theta$ is the angle between $B_0$ and the $[100]$ crystal axis \cite{Hasegawa1960, Wilson1961}.

Even if the electron wave function were entirely confined in one valley, strain can cause a change in $g$-factor by shifting the nearby energy bands that determine $g$ \cite{Wilson1961, Roth1960}. This "one-valley" mechanism yields a spin-lattice relaxation rate of the form:
 \begin{equation} \label{eq:fullT1b}
\begin{aligned} 
T_{1,\rm ov}^{-1} & =   \frac{1}{20\pi}\left(\frac{M}{g_0}\right)^2\left(\frac{\Xi_u}{E_{12}}\right)^2 \left(\frac{1}{\rho \bar{v}_t^5}+\frac{2}{3\rho\bar{v}_l^5}\right)\\
&   \times\left(\frac{g\mu_{\rm B}B_0}{\hbar}\right)^4f_{\rm ov}(\theta)\cdot k_{\rm B} T\\
& =  K_4^{\rm ov}B^4T,
\end{aligned}
 \end{equation}
where $M=0.44$ is the matrix element of the one-valley g-factor shift and  $f_{\rm ov}(\theta)=\cos^4\theta(1+1/2\sin^4\theta)$ \cite{Roth1960}. 
Since the magnetic field in our experiment is aligned along the [110] direction, $\theta=45^\circ$, both the "valley repopulation" and the "one-valley" mechanisms provide a channel for spin relaxation. 

The spin relaxation rates in Eq. \eqref{eq:fullT1} and \eqref{eq:fullT1b} were derived in the high-temperature limit, $k_{\rm B}T \gg g\mu_{\rm B}B_0$, where both spontaneous and stimulated phonon emission take place. These are described by including a factor $(1+n_{\rm ph})$ in the rate calculation, where $n_{\rm ph}=1/[\exp(g\mu_{\rm B}B_0/k_{\rm B}T)-1]$ is the Bose occupation factor for phonons of energy matching the electron Zeeman energy. The factor $k_{\rm B}T$ in Eqs.~\eqref{eq:fullT1}, \eqref{eq:fullT1b} appears because $(1+n_{\rm ph})\approx k_{\rm B}T / g\mu_{\rm B}B_0$ in the high-$T$ limit.

The low-temperature limit of the spin relaxation rates, of relevance to experiments we present here, is obtained by replacing $k_{\rm B}T / g\mu_{\rm B}B_0$ with 1 in Eqs.~\eqref{eq:fullT1}, \eqref{eq:fullT1b}, which results in the well-known $T_1^{-1} \propto B^5$ dependence \cite{Morello2010, Zwanenburg2013}:
\begin{equation}
T_1^{-1}|_{\rm low-T}=K_4\frac{g\mu_B}{k_B} B_0^5 = K_5 B_0^5.
\end{equation}  

\subsection{Evanescent-wave Johnson noise} \label{sec:magneticnoise}

Another mechanism inducing electron spin relaxation is magnetic noise leaking from the aluminum gates in the vicinity of the electron. Quantum and thermal fluctuations of the electrical currents in the metal create electromagnetic fluctuations known as Johnson noise \cite{Johnson1928, Nyquist1928, Callen1951}. The Johnson noise leaks out of the metal into the insulator in form of evanescent waves when the photon modes in the metal are totally reflected at the metal-insulator interface (Fig.  \ref{fig:ewjn}a) \cite{Volokitin2007}. This effect is called evanescent-wave Johnson noise (EWJN) \cite{Henkel1999, Poudel2013, Premakumar2017} and is particularly strong near a metal interface. EWJN can cause spin relaxation at low temperatures because the evanescent waves constitute an electromagnetic reservoir that can absorb energy (Eq. \ref{eq:fgr}).

In the nanoscale MOS devices studied here, the main sources of EWJN are the metallic control gates (see Fig.~\ref{fig:device}a). At $\boldsymbol{r}$, the position of the donor, this noise is characterized by the power spectrum
\begin{align}
S_{ii}\left( \omega \right) & =\int_{-\infty }^{\infty }e^{i\omega
t}\left\langle B_{i}\left( \mathbf{r},t\right) B_{i}\left( \mathbf{r}%
,0\right) \right\rangle ~dt \notag\\
& =\left\langle B_{i}\left( \mathbf{r}\right) B_{i}\left( \boldsymbol{r}%
\right) \right\rangle _{\omega }.
\end{align}
Here $i=x,y,z$ is a Cartesian index and $B$ the magnetic component of the EWJN field. The angle brackets denote a thermal average over the quantum states of the system. The power spectrum determines $T_{1}$ according to the formula
\begin{equation}
\frac{1}{T_{1}}=\left( \frac{\mu _{B}}{\hbar }\right) ^{2}\left[
S_{xx}\left( \omega _{0}\right) +S_{yy}\left( \omega _{0}\right) \right] ,
\end{equation}%
for $\mathbf{B}_{0}$ in the $z$ direction. 

As will be shown below, the conditions of our experiment are such that firstly we can approximate the electromagnetic fields as quasi-static, since the vacuum photon wavelength is on the order of cm and exceeds the device dimensions. Secondly, we anticipate a local relation between the electric field and the electric displacement since the devices satisfy the inequalities $\ell \ll a \ll \delta ,$ where $a$ is any linear dimension of the metal pieces, 
 \begin{equation} \label{eq:meanfreepath}
 \ell=v_{\rm F}\frac{m_e}{ne^2}\sigma
 \end{equation}
 is the mean free path with $v_{\rm F}$ as the electron Fermi velocity, $m_e$ the electron mass, $n$ the electron density, and 
\begin{equation} \label{eq:skindepth}
\delta=\sqrt{2/\mu_0\mu_R\sigma\omega_0}
\end{equation}
is the skin depth with $\mu_0$ as the magnetic permeability constant and $\mu_R$ as the relative permeability.  $\mu_R=1$ for our device.

For this situation, it has been shown that \cite{Premakumar2017}
\begin{equation}\label{eq:magneticT1}
\frac{1}{T_{1}}=\frac{1}{\mathcal{L}}\frac{\mu _{B}^{2}\mu _{0}^{2}\sigma \omega
_{0}}{4\pi \hbar },
\end{equation}%
where $\mathcal{L}$ is a length that depends only on the geometry of the metallic elements of the device and the position of the qubit.  Its calculation can be rather involved and we will give estimates for our device in Sec. IV A.

\subsection{Charge noise}

Charge noise does not directly couple to the spin of the qubit.  However, when combined with spin-orbit coupling, it creates a fluctuating effective magnetic field that will contribute to $T_1$ \cite{Huang2014}. According to Ref. \cite{Huang2014}, if the frequency dependence of the charge noise power spectrum is proportional to $1/f^a$, then the field dependence of $1/T_1$ is $B_{0}^{2-a}$.  For $1/f$ noise ($a=1$) this would give $1/T_1 \sim B_0$.  While a $1/f$ charge noise spectrum has been observed between $10^{-2}$ and $3\times10^{5}$~Hz in Si-based devices \cite{Yoneda2018}, it is exceedingly unlikely that it would hold up to the $> 10^{10}$~Hz frequency range that is relevant for $1/T_1$. Indeed, a recent re-analysis of the data in Ref.~\cite{Yoneda2018} suggests that the noise spectrum changes from $1/f$ to $1/f^2$ for $f>2\times 10^{5}$~Hz \cite{Gungordu2018}.

In the MOS donors-based devices discussed in this work, the  noise spectrum became white for $f \geq 10$ kHz \cite{Muhonen2014}. This would give $1/T_1 \sim B^2_0$ if extended up to the electron Larmor frequency.

\section{Qubit system and measurement methods} \label{sec:qubit} 

\subsection{Qubit setup}
\label{setup}

\begin{figure}
\centering
\includegraphics[width=0.9\columnwidth]{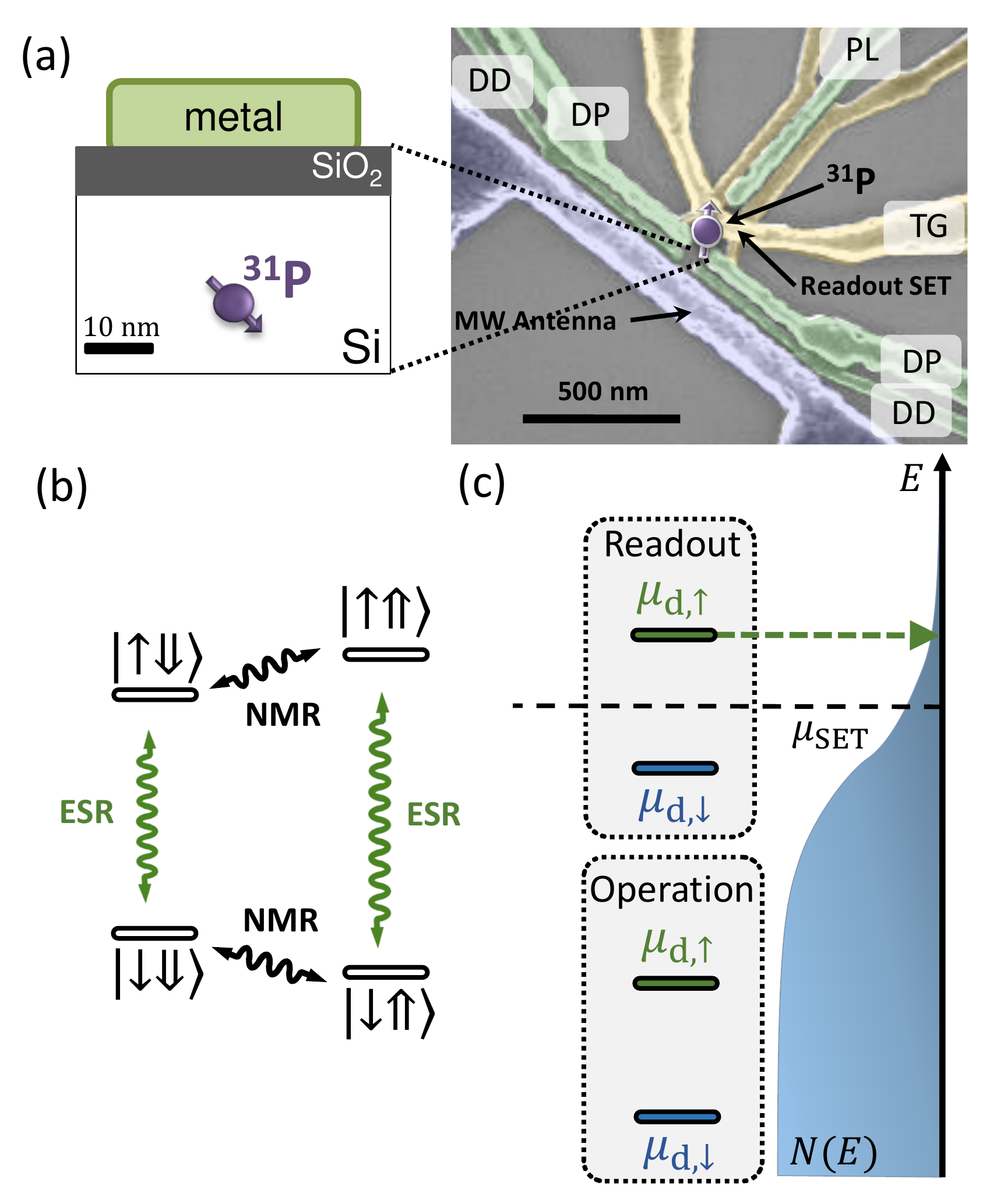}
\caption{(Color online) \textbf{Phosphorus donor qubit system.}
(a) Schematic of a phosphorus donor implanted in silicon, and a false-colored scanning electron micrograph of a device similar to the ones measured. Two independent gates control the donor potential (DD and DP, green) while a single-electron transistor (SET, yellow) determines the donor charge state, which is correlated to the electron spin state via a spin-dependent tunneling process. A plunger gate (PL, green) controls the electrochemical potential of the SET (and of a donor, if one is present in its vicinity). A broadband microwave antenna (purple) provides a magnetic drive for both the electron and the nuclear spins. (b) Energy level diagram of the electron-nuclear spin system, with electron spin resonance (ESR) and nuclear magnetic resonance (NMR) transitions indicated. (c) Schematic of the (electron-spin dependent) donor electrochemical potentials $\mu_{\rm d,\uparrow}, \mu_{\rm d,\downarrow}$ during readout and operation. The nuclear state is irrelevant in the readout process. For readout, the donor is tuned such that $\mu_{\rm d,\uparrow}-E_{\rm Z}/2=\mu_{\rm SET}=\mu_{\rm d, \downarrow}+E_{\rm Z}/2$ and $\ket{\uparrow}$ can tunnel out, leaving the donor ionized. The resulting positive donor charge shifts the SET tuning to a high-conductance point. In contrast, $\ket{\downarrow}$ stays confined, keeping the SET in Coulomb blockade. For operation, both spin states are well confined below the SET electrochemical potential with $\mu_{\rm d, \downarrow},\mu_{\rm d,\uparrow}\ll \mu_{\rm SET}$, so the electron cannot escape. This is the bias point for the device during the wait time for spin relaxation measurements. $N(E)$ is the density of states in the SET island. 
}
\label{fig:device}
\end{figure}

Our qubit system consists of a single electron spin confined by a phosphorus $^{31}$P donor, implanted in either natural silicon ($^{\rm nat}$Si) or isotopically-enriched $^{28}$Si with $800\,$ppm residual $^{29}$Si nuclei (Fig. \ref{fig:device}a) \cite{Itoh2014}. With the ion implantation parameters used for the devices described in the present work, each device contains typically $10-20$ donors in a $100 \times 100$~nm$^2$ window.

Aluminum gates, defined by electron beam lithography, control the electrostatic environment and allow selecting a specific donor for the measurements. Here, spin readout is obtained via spin-dependent tunneling into the island of a single-electron transistor (SET) \cite{Morello2010} kept at a low electron temperature ($T \approx 100$~mK). It is always possible to tune the gate voltages in such a way that one and only one donor has its electrochemical potential aligned with that of the SET island, while all other donors are either already ionized, or are kept far below the Fermi level (Fig. \ref{fig:device}c).

A DC-only (DD) and a pulsed (DP) gate above the donor control the donor potential. Additionally a plunger gate (PL) is used to manipulate the donor potential and the SET electrochemical potential $\mu_{\rm SET}$. In all devices from 2013 onward, a broadband microwave antenna \cite{Dehollain2013} is used for microwave and radio frequency pulses, allowing for full control over the electron \cite{Pla2012} and nuclear \cite{Pla2013} spins (Fig. \ref{fig:device}b). 

\subsection{\label{sec:Measurementprocedures} Measurement procedures}

\paragraph*{Donor control via a virtual gate.}
The spin readout process depends on the relative alignment of the (spin-dependent) donor electrochemical potentials $\mu_{\rm d,\uparrow}, \mu_{\rm d,\downarrow}$ with respect to the SET electrochemical potential $\mu_{\rm SET}$ \cite{Morello2009,Morello2010}. To simplify the analysis we define a virtual pulsed gate voltage $V_{\rm DV}$ by combining the effects of the voltage pulses on the SET plunger, $V_{\rm PL}^{\rm ac}$, and on the donor pulsed gates, $V_{\rm DP}^{\rm ac}$: 
\begin{equation}
V_{\rm DV}=\sqrt{(\beta V_{\rm PL}^{ac})^2+\left( V_{\rm DP}^{ac}\right)^2}.
\end{equation} 
These pulsed voltages are applied in addition to the DC voltages $V_{\rm PL}$ and $V_{\rm DP}$ chosen to select a specific donor to be near the readout condition.

\begin{figure}
\centering
\includegraphics[width=0.8\columnwidth]{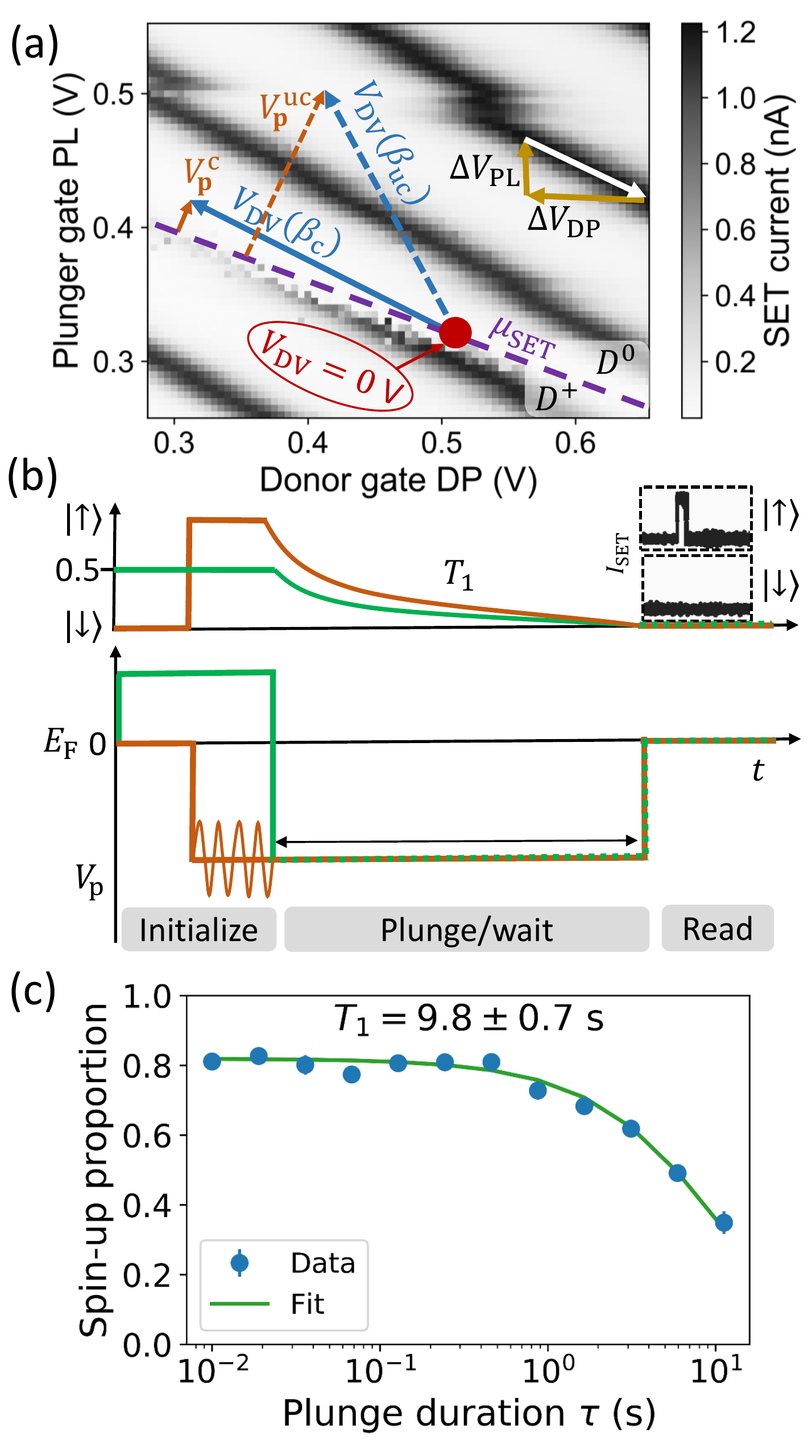}
\caption{(Color online) \textbf{Measurement of the electron spin relaxation rate. }
(a) Detail of the charge stability diagram [SET current as a function of plunger gate (PL) and donor gate (DP)] around the ionization point of the donor under study. The donor transition from ionized ($D^+$) to neutral ($D^0$) when $\mu_{\rm d}=\mu_{\rm SET}$ is indicated by the dashed purple line. The virtual gate voltage $V_{\rm DV}(\beta_{\rm c}=-0.51)$, indicated by the blue arrow, is determined by $\Delta V_{\rm PL}$ and $\Delta V_{\rm DP}$ (Eq. \ref{eq:betac}). $V_{\rm p}^{\rm c/uc}$ is the effective plunge voltage. (b) Schematic of the pulse sequence to measure the relaxation time $T_1$. For fields $B_0\leq1.5\,$T (orange line), the spin is determinstically initialized to $\ket{\downarrow}$ and then inverted with an adiabatic ESR pulse. For $B_0>1.5\,$T (green line) an electron with a random spin state is initially loaded. Then the donor is plunged for time $\tau$, until the spin state is determined by spin-dependent tunneling with the SET. (c) Example of a $T_1$ measurement at $B_0=1\,$T. The relaxation time $T_1=9.8\pm0.7\,$s is extracted using a least-square exponential fit (Eq. \ref{eq:Pup}), with $C_{\rm offset}=0.10$ determined in a separate experiment.
}
\label{fig:t1example}
\end{figure}

The factor $\beta$ determines the way in which we choose to shift $\mu_{\rm d}$ and $\mu_{\rm SET}$. We typically choose ``compensated pulses", i.e. keep $\mu_{\rm SET}$ fixed while moving $\mu_{\rm d}$ by using $V_{\rm PL}$ to compensate for the effect of $V_{\rm DP}$ on $\mu_{\rm SET}$. We thus call $\beta_{\rm c}$ the slope of the Coulomb peaks in the charge stability diagram of the donor and plunger gates (Fig. \ref{fig:t1example}a), determined by the ratio of capacitive couplings of gates PL and DP to the SET island:
\begin{equation}\label{eq:betac}
\begin{aligned}
\beta_{\rm c} & = & \Delta V_{\rm PL}/\Delta V_{\rm DP} \\
 & = & C_{\rm SET-PL}/C_{\rm SET-DP}.
\end{aligned}
\end{equation}
Any other value of $\beta$ corresponds to an uncompensated operation ($\beta_{\rm uc}$), i.e. one where $\mu_{\rm SET}$ varies during the pulsing. 

We also define the donor plunge voltage $V_{\rm p}^{\rm c}$ ($V_{\rm p}^{\rm uc}$) as the effective voltage that determines how far below $\mu_{\rm SET}$ the donor electrochemical potential $\mu_{\rm d}$ is plunged when operating compensated (uncompensated):
\begin{equation} \label{eq:V_p}
V_{\rm p}^{\rm c/uc}=V_{\rm DV}(\beta_{\rm c/uc})\sin\theta,
\end{equation} 
where $\theta=\angle \left[V_{\rm DV}(\beta_{\rm c/uc}), \mu_{\rm SET}\right]$.
Note that, for $V_{\rm p}^{\rm uc}$, the shift of $\mu_{\rm SET}$ caused by the uncompensated pulsing can result in a change in the electron number in the SET island. 

\paragraph*{Electron spin read out.}
The current through the SET, $I_{\rm SET}$, is used to determine the charge state of the donor which, in turn, correlates to the electron spin state in the presence of spin-dependent tunneling \cite{Martin2003,Elzerman2004,Morello2009,Morello2010}. The SET is biased in Coulomb blockade ($I_{\rm SET}\approx 0$) when the donor is in the neutral charge state. For spin readout, the donor and SET electrochemical potentials are tuned such that $\mu_{\rm d,\uparrow}-E_{\rm Z}/2=\mu_{\rm SET}=\mu_{\rm d, \downarrow}+E_{\rm Z}/2$. This ensures that the electron can only leave the donor and tunnel onto the SET if in state $\ket{\uparrow}$, leaving behind a positively charged donor which shifts the SET bias point and brings it to a high-conductance state ($I_{\rm SET} \approx 1$~nA). Coulomb blockade is restored when a $\ket{\downarrow}$ electron tunnels back onto the donor. Thus we observe a current spike whenever the electron was in state $\ket{\uparrow}$, while the current stays low if in state $\ket{\downarrow}$ (Fig. \ref{fig:device}c). This donor tuning is called``read level" and, in our definition, corresponds to $V_{\rm p}=V_{\rm DV}=0\,$V (Fig. \ref{fig:t1example}a).

\paragraph*{Electron spin initialization}
For $B_0 \leq 1.5$~T we prepare a $\ket{\uparrow}$ state in two steps. First we use the read level, $V_{\rm p}=0\,$V, to initialize $\ket{\downarrow}$. After a waiting time suitably longer than the electron tunnel-out time, a $\ket{\uparrow}$ will have escaped the donor and be replaced by a $\ket{\downarrow}$, while $\ket{\downarrow}$ will remain in place. Second, we invert the spin from $\ket{\downarrow}$ to $\ket{\uparrow}$ using an oscillating magnetic field $B_1$ whose frequency is adiabatically swept through the resonance \cite{Laucht2014b}.

For $B_0 > 1.5$~T the above method would require ESR frequencies higher than those available with our microwave source. We thus resort to a random electron initialization, obtained by ,loading the electron when $\mu_{\rm d, \downarrow},\mu_{\rm d,\uparrow}\ll \mu_{\rm SET}$. In this case both the $\ket{\uparrow}$ and $\ket{\downarrow}$ states are accessible and electron spin is prepared with roughly equal probability of the two.

\paragraph*{Spin relaxation measurement.}
The electron spin relaxation time $T_1$ is obtained by measuring the probability of finding the spin in the $\ket{\uparrow}$ state after a wait time $\tau$ has elapsed. To this end, we apply the pulse sequence illustrated in Fig. \ref{fig:t1example}b to the virtual gate DV. 

For $B_0 \leq 1.5$~T we prepare a $\ket{\uparrow}$ state while for $B_0 > 1.5$~T a random electron is initialized with roughly equal probability of $\ket{\uparrow}$ and $\ket{\downarrow}$ (see paragraph \textit{Electron spin initialization}). 

Next, we plunge the donor electrochemical potential far below $\mu_{\rm SET}$ with a voltage pulse of amplitude $V_{\rm p}$ and duration $\tau$. This ensures that the previously initialized electron spin cannot escape the donor (see, however, Sect.~\ref{sec:cotunnelling}). Finally, a single shot-spin readout is performed at $V_{\rm p}=0\,$V. 

We repeat this sequence $30$ times to determine the spin-up fraction $P_{\uparrow}$ after each wait time $\tau$. The measurement of $P_{\uparrow}(\tau)$ is repeated multiple times to check for consistency, which can be occasionally disrupted by drifts and jumps in the electrostatic environment.

$T_1$ is extracted by performing a least-square fit to $P_{\uparrow}(\tau)$ with the exponential decay:
\begin{equation}
P_\uparrow(\tau)=C_{\rm init}e^{-\tau/T_1}+C_{\rm offset}, \label{eq:Pup}
\end{equation}
where $C_{\rm init}$ is the initial spin-up proportion and $C_{\rm offset}$ the offset at $\tau \rightarrow \infty$ created by erroneous spin-up counts, caused e.g. by tunnel-out events of $\ket{\downarrow}$ spins into states made available in the electron reservoir by thermal excitations, or by noise spikes counted as $\ket{\uparrow}$ spins. $C_{\rm init}$ and $T_1$ are free fitting parameters, whereas $C_{\rm offset}$ is determined separately by measuring the spin-up proportion after $\ket{\downarrow}$ has been initialized, and is fixed at that value in the fit. 

As an example, Fig. \ref{fig:t1example}c shows set of $P_{\uparrow}(\tau)$ that, fitted to Eq.~\eqref{eq:Pup}, yielded the longest measured relaxation time at $B_0=1$T, $T_1=9.8\pm0.7\,$s.

\section{Relaxation rate dependence on external magnetic field} \label{sec:extB}

\begin{figure*}
\centering
\includegraphics[width=1.55\columnwidth]{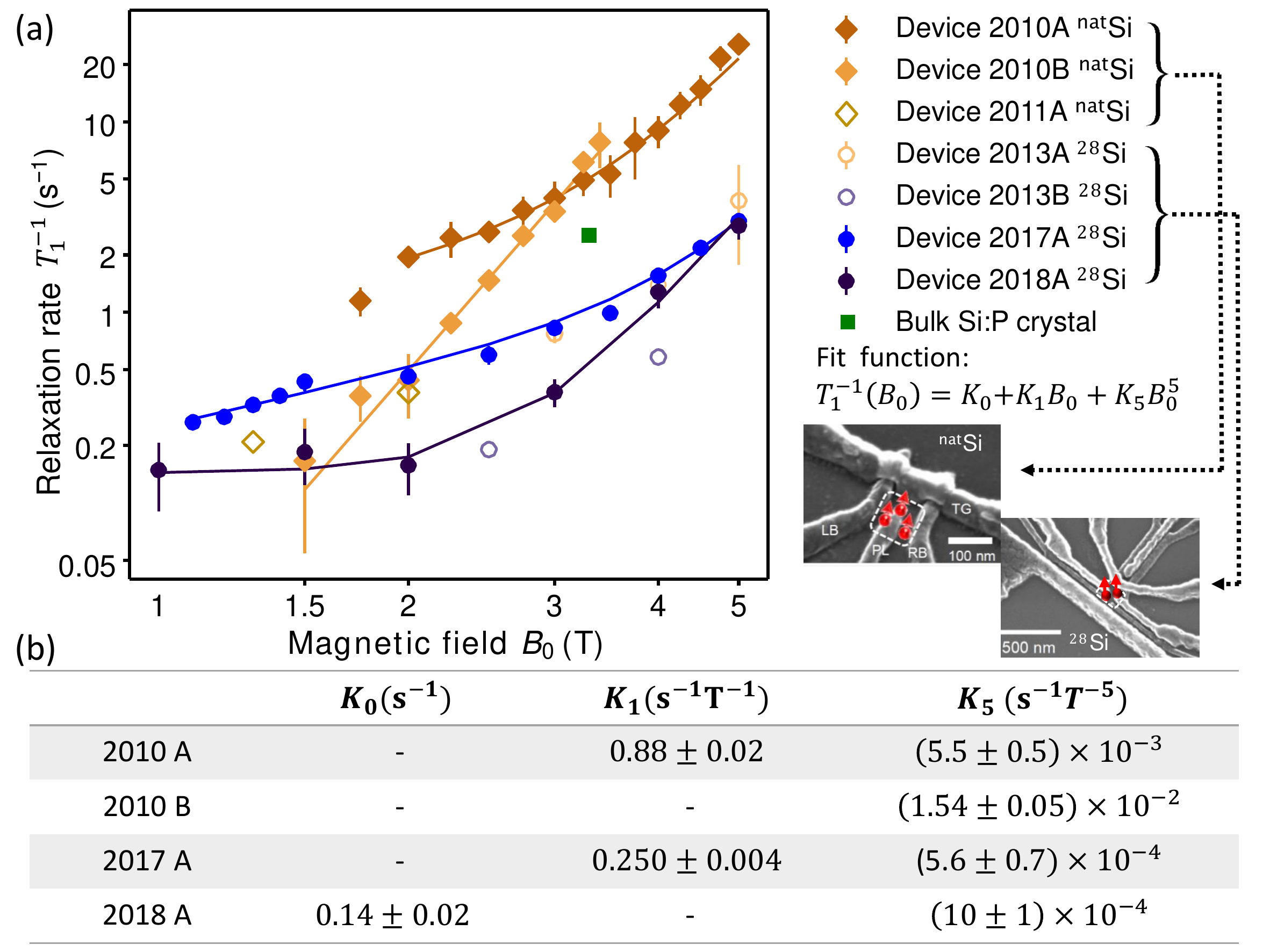}
\caption{(Color online) \textbf{Relaxation rate as a function of external magnetic field.}
(a) Measurements of the electron relaxation rate $1/T_1$ as a function of external magnetic field $B_0$ for different samples. Devices 2010A and 2010B are republished from Ref. \cite{Morello2010} and fabricated on $^{\rm nat}$Si, same as device 2011A (diamonds). Devices 2013A, 2013B, 2017A and 2018A have been fabricated on isotopically-enriched Si$^{28}$ epilayers (dots). For reference, a data point measured on a bulk Si:P crystal at $T<5\,$K is shown (green square, J. J. L. Morton, personal communication). For devices 2010A, 2010B, 2017A and 2018A, polynomials of the form $T_1^{-1}(B_0)=K_0+K_1B_0+K_5B_0^5$ have been fitted to the relaxation rate. The insets show the respective device designs. (b) Fitting parameters for the different samples. A dash indicates that the parameter was fixed at $K_i=0$.
}
\label{fig:magnetic field dependence}
\end{figure*}

The dependence of the electron relaxation rate $T_1^{-1}$ on the strength of the external magnetic field $B_0$ gives insight into the mechanisms that lead to the relaxation itself. Fig.  \ref{fig:magnetic field dependence} (a) shows sets of relaxation rates as a function of $B_0$ for seven different donor qubit devices, fabricated and measured in our laboratory between 2010 and 2018. Devices 2010A, 2010B (described in Ref.~\onlinecite{Morello2010}) and 2011A were fabricated on $^{\rm nat}$Si. Devices 2013A, 2013B, 2017A (described in Ref.~\onlinecite{Muhonen2014}), 2018A were fabricated on enriched $^{28}$Si. We fit the relaxation rate of devices 2010A, 2010B, 2017A and 2018A with a polynomial function of the form: 
\begin{equation}
T_1^{-1}(B_0)=K_0+K_1B_0+K_5B_0^5,
\end{equation}
with the results displayed in Table \ref{fig:magnetic field dependence}b (a dash indicates that the parameter was fixed at $K_n=0$). 

The prefactor $K_5$ describing the phonon-induced relaxation rate $\propto B_0^5$ at high magnetic fields varies significantly between the different devices (see Sec.~\ref{sec:phonon}).
Furthermore, all fitted devices show a deviation from $T_1^{-1}\propto B_0^5$ at magnetic fields $B_0\lesssim3\,$T, except for device 2010B: devices 2010A and 2017A follow $T_1^{-1}\propto B_0$, while device 2018A shows a $T_1^{-1}\sim \rm{const.}$ behavior at low field.

These deviations from bulk-like relaxation behaviors unveil details of the interaction betwen the donor electron spin and its environment in the MOS nanostructures under study.

\subsection{\label{sec:ewjn}Relaxation induced by Evanescent-wave Johnson noise}

\begin{figure}
\centering
\includegraphics[width=\columnwidth]{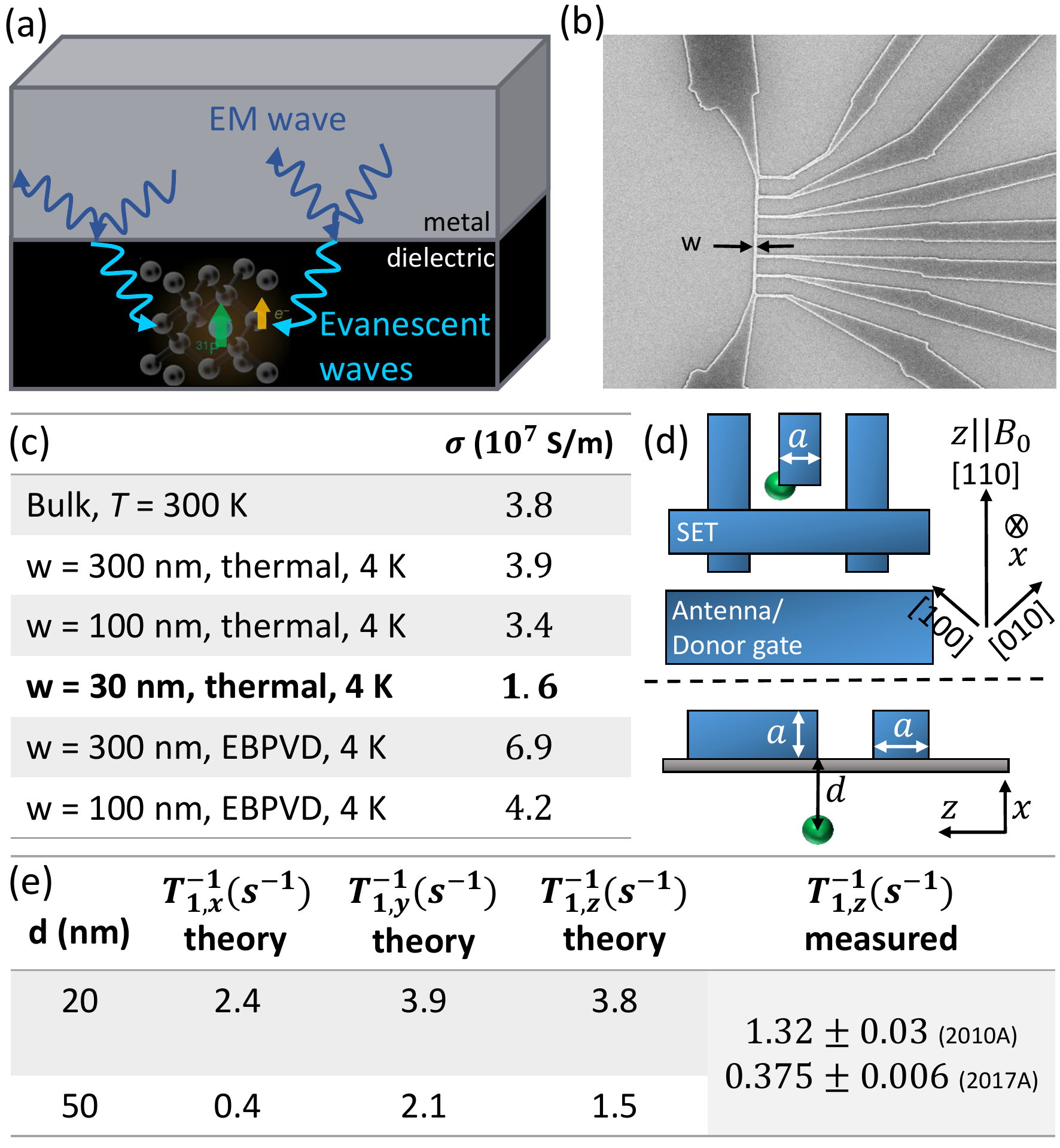}
\caption{(Color online) \textbf{Evanescent wave Johnson noise.}
(a) Schematic of the origin of EWJN in our qubit devices. (b) Scanning electron micrograph of a Hall bar structure to measure the conductivity $\sigma$ of the Al gates. The width is $w=30$~nm in the depicted device. (c) Conductance of 50~nm thick Al metal, measured at temperature $T=4$~K with Hall bar structures as in (b), for different feature widths $w$, formed either by thermal evaporation or electron-beam physical vapour deposition (EBPVD). Room-temperature bulk value for comparison. (d) Device layout showing the donor position, the external magnetic field and the crystal orientation. $a$ is the metallic gate dimension, $d$ is the distance between the donor and the metal gates. (e) Relaxation rates predicted by the EWJN theory using $\sigma=1.6\times10^7$~S/m, $B_0=1.5$~T, $a=50$~nm and $d=20$ or $50$~nm, compared to two measured values.  
}
\label{fig:ewjn}
\end{figure}

In our metal-oxide-semiconductor devices, the electrostatic  gates, SET, and microwave antennas are all potential sources of EWJN. 

Replacing the qubit Larmor frequency with $\omega_0=g\mu_B B_0/\hbar$ in Eq. \eqref{eq:magneticT1} yields:
\begin{align} \label{eq:ewjnsimple}
T_1^{-1} & = \frac{\mu_B^2\mu_0^2\sigma g B_0}{4\pi\hbar^2}\frac{1}{\mathcal{L}}\notag\\
& = K_1 B_0.
\end{align}
The most important point about this formula is that no other plausible spin relaxation mechanism gives a rate proportional to $B_0$. Linearity of $T_1^{-1}$ in $B_0$ is thus a convincing signature of EWJN.

For the validity of the analysis that follows, the value of the electrical conductance $\sigma$ of the aluminum structures is very important.  $\sigma$ determines the characteristic length scales $\ell$ (mean free path) and $\delta$ (skin depth) and the resulting magnitude of the relaxation.  We extracted $\sigma$ from 4-point measurements on Hall bar structures (Fig.  \ref{fig:ewjn}b) with feature sizes varying from $300\,$nm to $30\,$nm. We tested aluminum layers formed both via thermal evaporation and electron beam physical vapour deposition (EBPVD), but all devices on which spin relaxation was measured and reported in Fig.  \ref{fig:magnetic field dependence} were fabricated using thermal evaporation.   

We find that the conductivity drops with reduced feature size but only up to a factor of 2 (Tab. \ref{fig:ewjn}c), which is consistent with a grain size of approximately $20\,$nm, i.e. comparable but still smaller than the width and thickness of the fabricated gates. We base the calculations below on the value $\sigma=1.6\times 10^7\,$S/m obtained for the $30\,$nm feature size, which corresponds to the smallest gate dimensions used in donor devices studied in this paper. This conductance results in a skin depth  $\delta(B_0=1\,\rm{T})=752\,$nm (Eq. \ref{eq:skindepth}) and a mean free path  $\ell=6.3\,$nm (Eq. \ref{eq:meanfreepath}) with $\mu_R=1$, $n=18\times 10^{28}\,\rm{m}^{-3}$ and $v_{\rm F}=2\times 10^2\,$m/s \cite{Ashcroft1976}. This shows that $\ell$ is always smaller than even the smallest feature sizes in our devices, placing the conduction electrons in the aluminum gates in the diffusive regime. 

EWJN depends on the gate geometry through the geometric factor $\mathcal{L}$ (Eq. \ref{eq:ewjnsimple}). $\mathcal{L}$ can be calculated analytically for different cases: half spaces and spheres.  The electron spin effectively sees a metallic half space when its distance to the gates $d$ is much smaller than the gate lateral dimensions $a$. When the spin is further away from a finger gate or an antenna ($d\gg a$, Fig. \ref{fig:ewjn}d), it sees approximately a conducting cylinder. Since our devices have $d\approx10-20\,\rm{nm}$ and $a\approx30-80\,$nm, we employ an interpolation between both cases in form of
\begin{equation}
1/T_{1i} = 1/\left[T_{1i}(\mathcal{L}_{\rm hs})+T_{1i}(\mathcal{L}_{\rm cyl})\right],
\end{equation}
where $i$ indicates the direction $x,y$ or $z$ of the applied field $B_0$. We model an antenna or finger gate as a string of spherical beads.  
The final relaxation rate follows as 
\begin{subequations}
\begin{equation}
T_{1,x}^{-1}=\frac{\mu_B^2\mu_0^2\sigma\omega_0}{32\pi\hbar d} \left(1+\frac{256d^4}{15\pi a^4}\right)^{-1},
\end{equation}
\begin{equation}
T_{1,y}^{-1}=\frac{3\mu_B^2\mu_0^2\sigma\omega_0}{64\pi\hbar d} \left(1+\frac{256d^4}{91\pi a^4}\right)^{-1},
\end{equation}
\begin{equation}
T_{1,z}^{-1}=\frac{3\mu_B^2\mu_0^2\sigma\omega_0}{64\pi\hbar d}\left(1+\frac{256d^4}{47\pi a^4}\right)^{-1}.
\end{equation}
\end{subequations}
Using the measured conductivity (Tab. \ref{fig:ewjn}c), the predicted relaxation rate due to EWJN for $B_0=1.5$~T applied in the $z$-direction (in the plane of the device) is $T_{1z}^{-1}\approx 4\,\rm{s}^{-1}$, for a donor depth $d=20\,$nm and aluminum gates of width $a=50\,$nm (Tab. \ref{fig:ewjn}e). This prediction is close to the measured value of $T_1^{-1}=1.3\,$s$^{-1}$ in device 2010A, while it overestimates $T_1^{-1}$ by around one order of magnitude for device 2017A. Neither device 2010B nor device 2018A exhibit a $T_1^{-1}\propto B_0$ behaviour within the measured range of magnetic fields.

This order of magnitude agreement between theory and experiment can be considered satisfactory, in light of the many experimental parameters that are only approximately known, such as the donor depth $d$, as well as the lateral position of the donor with respect to the gates (the devices that show no evidence of $1/T_1 \propto B_0$ could have the donor underneath the gaps between the gates, for example).

Table~\ref{fig:ewjn}e shows the predicted anisotropy of $1/T_1$ as a function of the direction of $B_0$. In the future, such anisotropy of the EWJN contribution could provide a further test of the theory, if $1/T_1$ were measured as a function of field direction using a 3D vector magnet.

\subsection{Phonon-induced relaxation: effects of lattice strain} \label{sec:phonon}

The phonon-induced electron spin relaxation strongly depends on the crystalline environment of the donor. We observed nearly two orders of magnitude variation in the prefactor $K_5$ of the term $T_1^{-1}\sim B_0^5$ (Fig.~\ref{fig:magnetic field dependence}). We tentatively attribute this variability to the variation of local strain in the devices. Strain in MOS devices arises due to the different thermal expansion coefficients of aluminum and silicon \cite{Thorbeck2015}. The donors are quite close to the Al gates, and the presence of strain has been documented in several experiments, especially for its impact on the hyperfine coupling $A$ \cite{Laucht2015,Pla2018,Mansir2018}. 

As shown in Eq.~\eqref{eq:energyshift}, the valley energies shift with strain. This leads to a lowering in energy of the $E$ excited  states [Eq.~\eqref{eq:valleys}e,f], i.e. to a reduction of the valley-orbit splitting $E_{12}$ \cite{Wilson1961, Tahan2002}, which would suggest that the spin relaxation becomes faster with strain [see Eqs.~\eqref{eq:fullT1},\eqref{eq:fullT1b}]. However, for large compressive strain in the $z$-direction the lowest-energy valley-orbit states become symmetric and antisymmetric combinations of the $\pm z$ valleys. This causes the overlap matrix element $D_{\bm{r}}$ [Eq.~ \eqref{eq:couplMatrix}] to become vanishingly small \cite{Tahan2002}. The decrease of $D_{\bm{r}}$ caused by the change in valley composition greatly outweighs the increase of $1/E_{12}$, resulting in an overall reduction $T_1^{-1}$, according to Eq.~\eqref{eq:ephH}.  
 
Device 2017A was also the subject of the experiments by Laucht \textit{et. al.} \cite{Laucht2015}. In that work, the analysis of the hyperfine shift yielded $s_{xy}\approx-0.1\%$ in-plane compressive strain. This device exhibits the slowest phonon-induced relaxation (lowest $K_5$) among all tested and, significantly, the strongest deviation from the bulk value of the hyperfine coupling ($A\approx 97$~MHz). 

In Device 2018A we measured $A\approx 115$~MHz from which, using the atomistic tight binding simulations from Fig. S6 in Ref. \cite{Laucht2015}, we estimate a strain $s_{xy}\approx-0.05\%$. This lower value of the strain is consistent with the faster spin-phonon relaxation observed in this device ($K_5\approx 10 \times 10^{-4}$~s$^{-1}$T$^{-5}$, compared to $K_5\approx 5 \times 10^{-4}$~s$^{-1}$T$^{-5}$ in Device 2017A). 

The highest value of $K_5 \approx 1.5\times10^{-2}$ was found in Device 2010B. That device did not have a microwave antenna, so the hyperfine coupling could not be measured. Interestingly, $1/T_1$ in Device 2010B coincides with the relaxation rate measured in an all-epitaxial single-donor device fabricated via STM hydrogen lithography \cite{Watson2015}. The STM device is likely to exhibit very little strain, since the donor is deeply embedded in the silicon crystal and no metal gates are present in its vicinity. These findings suggest that device 2010B contained a donor implanted deeper than usual, far away from the aluminium gates, and therefore subjected to a reduced amount of strain. The deep location of the donor would also explain the absence of EWJN-induced relaxation in this device, which followed $1/T_1 \propto B_0^5$ down to the lowest field.  

Observing the trend of phonon-induced relaxation across all devices, one might notice that devices fabricated on $^{28}$Si epilayers appear to always have longer $T_1$ than those on natural silicon. This could be due to some built-in strain in the epilayers.

\subsection{Other relaxation processes}

One device, 2018A, exhibits a field-independent relaxation rate for $B_0 < 2$~T. In Ref.~\onlinecite{Morello2010}, the relaxation rate of Device 2010A was also interpreted as a combination of $1/T_1\propto B_0^5$ and $1/T_1 = const.$ (the data point at $B_0=1.75$~T was thought to be an outlier), and a quantitative model was developed to justify the constant contribution. Since our devices contain on average $10-20$ donors in a $100\times 100$~nm$^2$ region, we analyzed the rate at which a spin excitation on the donor under measurement can diffuse to nearby donors by means of magnetic dipole-dipole interactions. The flip-flop rate $\Gamma_{\rm ff}$ between a pair of donors can be expressed as \cite{Morello2010}:
\begin{equation}
    \Gamma_{\rm ff} \approx \frac{\pi}{2\langle \Delta \omega_I \rangle}M_{\rm ff}(\theta,d),
    \label{Eq:Gamma_ff}
\end{equation}
where $\langle \Delta \omega_I \rangle$ is the half-width of each electron spin resonance as caused by the Overhauser field from the $^{29}$Si nuclei, and $M_{\rm ff}$ is the flip-flop matrix element in the magnetic dipolar coupling Hamiltonian, which depends on the angle $\theta$ and the distance $d$ between the spins. This model yields $\Gamma_{\rm ff} \approx 2$~s$^{-1}$ using $d=24$~nm and taking $\langle \Delta \omega_I \rangle/\hbar \approx 3.5$~MHz \cite{Pla2012} as the typical value of Overhauser field broadening in $^{\rm nat}$Si.

It is immediately clear from Eq.~\eqref{Eq:Gamma_ff} that this model would yield implausible results when applied to the $^{28}$Si enriched samples, where $\langle \Delta \omega_I \rangle/\hbar \approx 1$~kHz is three orders of magnitude smaller than in $^{\rm nat}$Si \cite{Muhonen2014}. This is because Eq.~\eqref{Eq:Gamma_ff} assumes that the donors have the same hyperfine coupling $A$ and the same $g$-factor, and their resonance frequencies are detuned solely by Overhauser fields. We now know that this assumption is, in general, unlikely to hold: we have observed hyperfine couplings ranging from $97$ to $116$~MHz in various devices, with the spread arising from different local electric fields and strain \cite{Laucht2015}. Including the effect of locally different $A$ and $g$ for different donors within the same device would result in a near-complete suppression of the (energy-conserving) flip-flop processes. Therefore, we do not believe that this mechanism can be responsible for the field-independent relaxation rate observed in Device 2018A.

\begin{figure}
\centering
\includegraphics[width=\columnwidth]{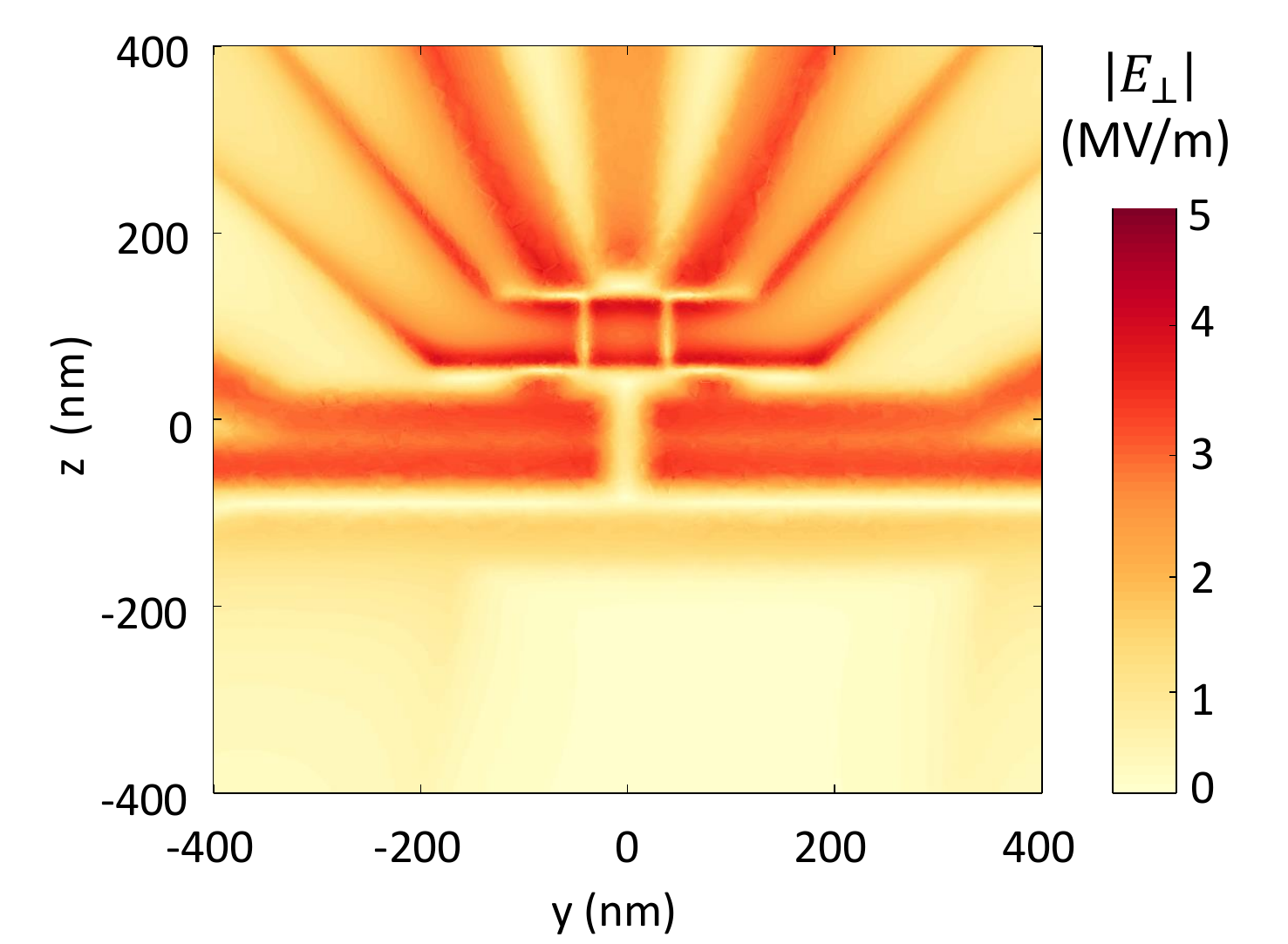}
\caption{(Color online) \textbf{Perpendicular electric field at the typical donor depth.}
Absolute value of the electric field component $|E_\perp|$ perpendicular to the magnetic field $B_0$ applied along the $z$-axis, calculated using the COMSOL finite-elements electrostatic package. We show values $10\,$nm under the Si/SiO$_2$ interface, a typical donor implantation depth. The model consists of a $2\,\mu$m$\times2\,\mu$m$\times2\,\mu$m silicon substrate grounded at the bottom. On top of a $8\,$nm SiO$_2$ layer, we define the aluminum gates, which are coated by $2\,$nm of Al$_2$O$_3$. We assume typical voltages, i.e. $0.4\,$V to the barrier, plunger and donor gates, $2\,$V to the top gate, and we set the potential of the microwave antenna at ground. To model the 2DEG under the top gate, we ground the Si/SiO$_2$ interface in the relevant regions. 
}
\label{fig:eperp}
\end{figure}

Another relaxation mechanism, recently discovered in STM-fabricated donor devices \cite{Weber2018}, is a spin-orbit coupling (SOC) induced by the presence of an electric field $E_{\perp}$ perpendicular to the external magnetic field $B_0$. In our devices, the direction and strength of the electric field at the donor can vary significantly, depending on where exactly the donor is located with respect to the gates (Fig. \ref{fig:eperp}). An electric field component $E_{\perp}$ perpendicular to $B_0$ should, in general, be expected. This mechanism would mediate an additional spin-phonon relaxation channel on top of the bulk-like valley repopulation and one-valley relaxation, resulting in values of $K_5$ higher than in the bulk. Instead, in all devices except 2010A and 2010B, we found $K_5$ to be lower than the bulk value. This does not mean that this SOC mechanism does not exist in our devices, but it indicates that, in almost all cases, its contribution is less significant than the suppression of the relaxation rate caused by local strain.

\section{\label{sec:cotunnelling} Tunneling effects}

\begin{figure}
\centering
\includegraphics[width=0.8\columnwidth]{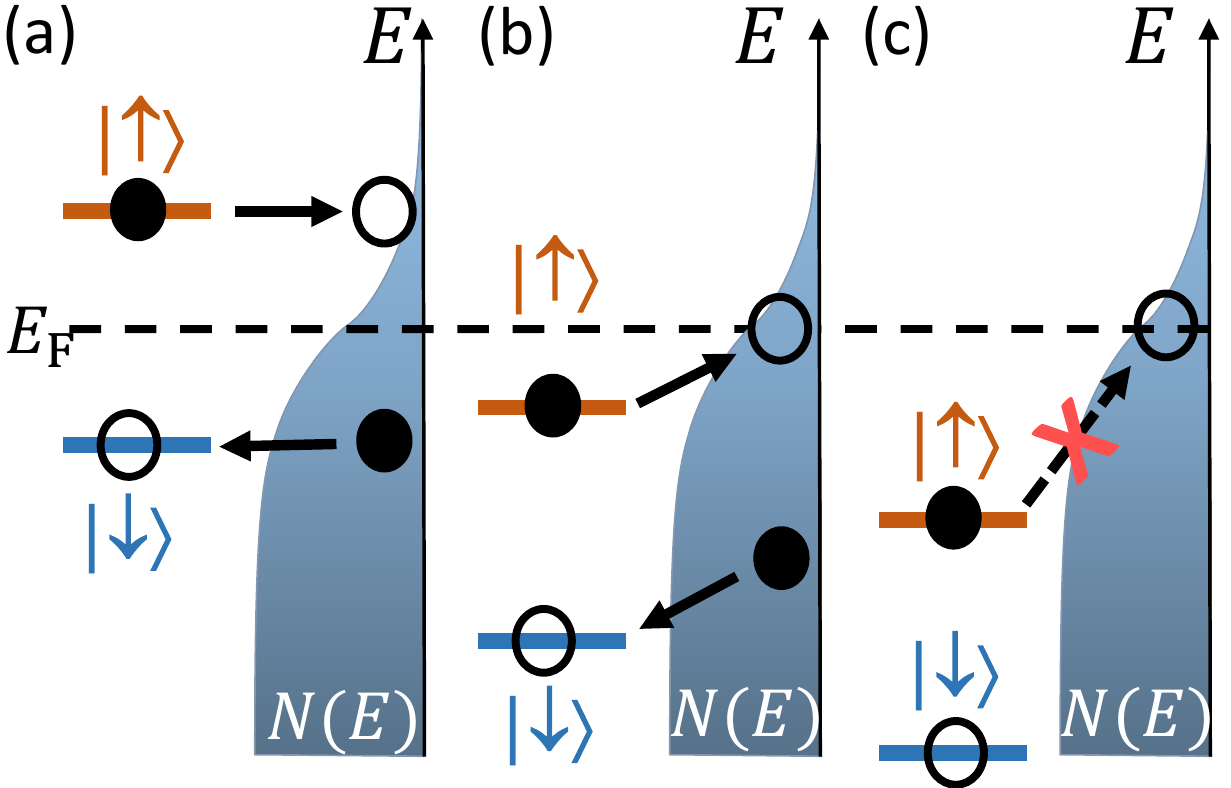}
\caption{(Color online) \textbf{Spin relaxation through quantum tunneling.} 
(a) Spin relaxation via direct tunneling of $\ket{\uparrow}$ from the donor into the SET reservoir while $\mu_{\rm d, \uparrow} \gtrsim \mu_{\rm SET}$; the $\ket{\uparrow}$ electron is then replaced by $\ket{\downarrow}$. (b) Spin relaxation via co-tunneling of $\ket{\uparrow}$ into a virtual free state in the SET while $\ket{\downarrow}$ tunnels onto the donor. (c) For $\mu_{\rm d}\ll \mu_{\rm SET}$ all tunnel processes are suppressed. 
}
\label{fig:tunnelsketch}
\end{figure}

\begin{figure*}
\centering
\includegraphics[width=1.8\columnwidth]{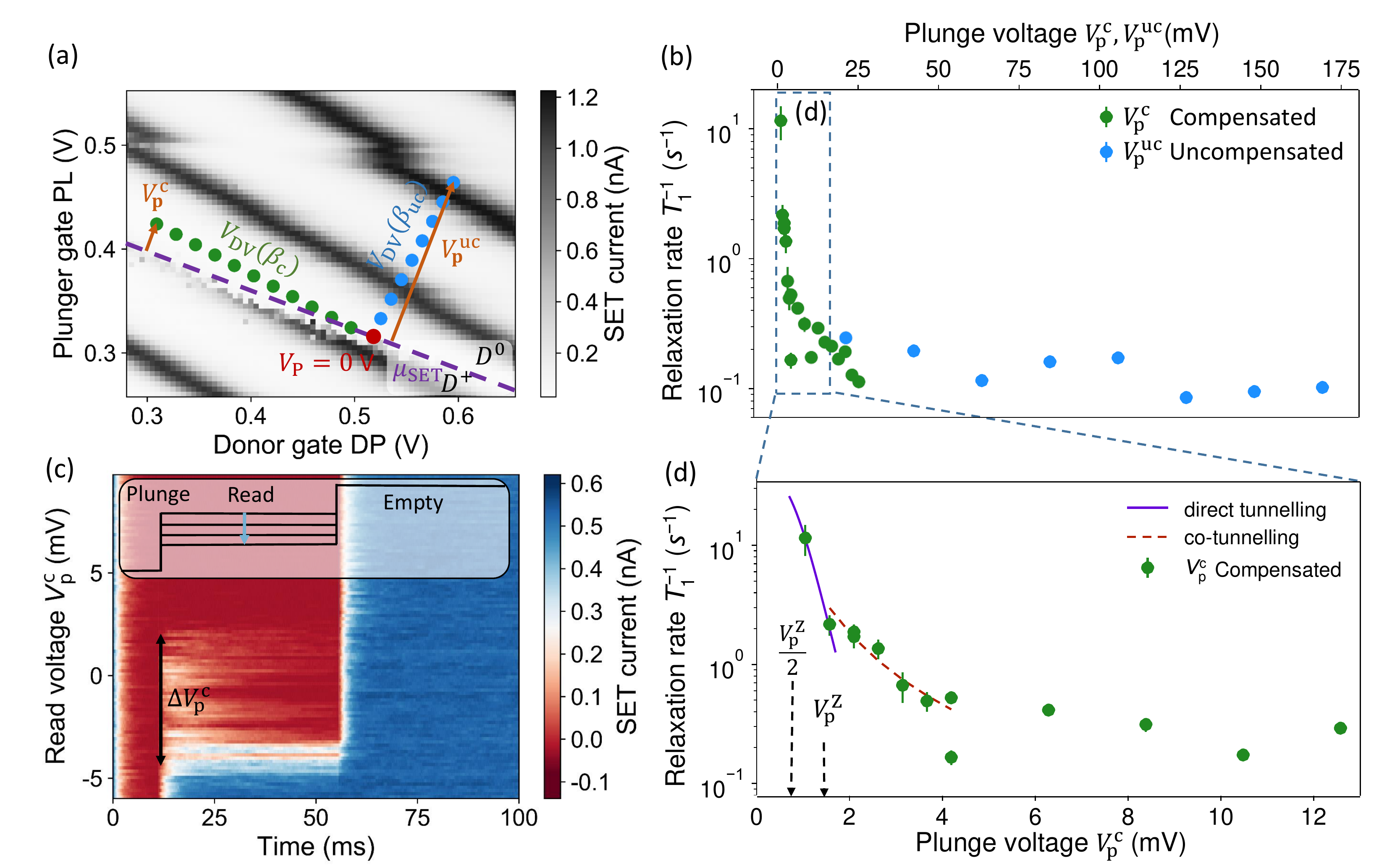}
\caption{(Color online) \textbf{Relaxation rate as a function of plunge voltage. }
(a) Charge stability diagram of the donor-SET system, obtained b monitoring the SET current as a function of plunger gate (PL) and donor gate (DP) voltages. The donor transition from ionized ($D^+$) to neutral ($D^0$) when $\mu_{\rm d}=\mu_{\rm SET}$ is indicated by the dashed purple line. Virtual gate voltage $V_{\rm DV}(\beta_{\rm c}=-0.51)$ (compensated plunging, green) and $V_{\rm DV}(\beta_{\rm uc}=1.81)$ (uncompensated plunging, blue) are indicated. The corresponding effective plunge voltage is $V_{\rm p}^{\rm c/uc}$ (orange arrows). (b) Relaxation rates as a function of $V_{\rm p}^{\rm c/uc}$. The dotted region indicates the area expanded in panel (d). (c) SET current as a function of the read level voltage $V_{\rm p}^{\rm c}$, resulting in a spin tail of length $\Delta V_{\rm p}^{\rm c/uc}=7.2$~mV at $B_0=5$~T. The inset shows the applied pulse sequence. (d) Zoomed-in plot for low plunge voltages with voltage $V_{\rm p}^{\rm Z}$ ($V_{\rm p}^{\rm Z}/2$) corresponding to the Zeeman energy $E_{\rm Z}$ ($E_{\rm Z}/2$) marked. The direct tunnelling process is described by Eq. \eqref{eq:directt} (purple line), using a bare tunnel rate $\Gamma_0 = 50$~s$^{-1}$. The red dotted line is an attempt to fit the region of slow decrease of $1/T_1$ to Eq.~\ref{eq:cot} which describes a second-order co-tunnelling process. However, a good fit to the data requires using a bare tunnel rate $\Gamma_0 = 1.4 \times 10^6$~ s$^{-1}$, in severe discrepancy with the value used for the direct tunneling fit. 
}
\label{fig:plungedependence}
\end{figure*}

The experiments described in this work rely upon switching between a ``plunge/wait" phase, during which the electron remains bound to the donor while its spin is allowed to relax, and a ``read'' phase, during which electron tunneling between the donor and the SET island is used to measure the spin state (Fig.~\ref{fig:device}c). Here we discuss the impact on the measurement results of the possibility that the electron tunnels out of the donor during the plunge/wait phase. 

To describe the rate of first-order tunneling between donor and SET island, we first define $\alpha_{\rm p}$ as the lever arm of the gate voltages to the donor, which determines the shift in $\mu_{\rm d}$ induced by the effective donor plunge $V_{\rm p}$ [see Eq.~\eqref{eq:V_p}]:
\begin{subequations} \label{eq:leverarm}
\begin{align}
\Delta \mu_{\rm d} = -e\alpha_{\rm p} V_{\rm p}\\
\alpha_{\rm p} =  \frac{\beta_{\rm c}C_{\rm d-PL}+C_{\rm d-DP}}{C_{\sum}}
\end{align}
\end{subequations} 
where $C_{\rm d-PL}$ and $C_{\rm d-DP}$ are the capacitances between the donor and the plunger gate PL and the donor and the donor gate DP, respectively, and $C_{\sum}$ is the total capacitance of all gates to the donor. In the presence of a magnetic field we define the donor electrochemical potential as the average of the $\ket{\downarrow}$ and $\ket{\uparrow}$ levels:
\begin{equation}
    \mu_{\rm d} = \frac{\mu_{\rm d,\downarrow} + \mu_{\rm d,\uparrow}}{2}.
\end{equation}

With this definition, the direct (first-order) tunnel-out rate of the $\ket{\uparrow}$ electron at electrochemical potential $\mu_{\rm d,\uparrow}$ can be written as \cite{Golovach2004, MacLean2007}: 
\begin{equation}\label{eq:directt}
\Gamma_{\rm DT} \approx \Gamma_0\cdot [1-f(V_{\rm p}, T_e)],
\end{equation}
where
\begin{equation}
    f(V_{\rm p},T_e)=\frac{1}{\left(1+\exp\frac{-e\alpha _{\rm p}V_{\rm p}+E_Z /2}{k_B T_e}\right)}
\end{equation}
is the Fermi function, $T_e$ is the electron temperature of the SET island and the term $-e\alpha _{\rm p}V_{\rm p}+E_Z /2 = \mu_{\rm d,\uparrow} - \mu_{\rm SET}$ describes the energy detuning between the $\ket{\uparrow}$ state and the SET electrochemical potential at a plunge voltage $V_{\rm p}$, with gate lever arm $\alpha_{\rm p}$. Since $k_B T_e \ll E_Z$ in our experiments, $\Gamma_0$ effectively represents the bare $\ket{\uparrow}$ tunnel-out rate at the ``read" position. For simplicity, we assumed that $\Gamma_0$ remains independent of $V_{\rm p}$ within the small voltage range used in the experiment. 

This direct tunnel process results in an apparent electron spin relaxation, when $\ket{\uparrow}$ tunnels from donor to SET and is replaced by a different electron in state $\ket{\downarrow}$ (Fig. \ref{fig:tunnelsketch}a). The spin relaxation rate is therefore similar (although not identical \cite{Otsuka2017}) to the charge tunneling rate. In this work, we have used the first-order tunneling process to deliberately initialize the spin in the $\ket{\downarrow}$ state for the experiments at $B_0 \leq 1.5$~T. Direct tunneling is exponentially suppressed with the energy difference between $\mu_{\rm d}$ and $\mu_{\rm SET}$ and is only expected as long as $\ket{\uparrow}$ is aligned with available free states in the SET island, above or just below $\mu_{\rm SET}$. 

Even if no free states are available for first-order tunneling, the $\ket{\uparrow}$ electron can relax via a second-order tunneling process. If an empty state at energy $E > \mu_{\rm d,\uparrow}$ is available in the electron reservoir, the $\ket{\uparrow}$ donor electron can virtually occupy such state for a time $t_{\rm H}\sim \hbar/(E-\mu_{\rm d,\uparrow})$ given by the Heisenberg uncertainty principle. During this time, another electron coming from the reservoir can occupy the donor state. This process can be inelastic if the original $\ket{\uparrow}$ electron is replaced by a $\ket{\downarrow}$ electron from the reservoir (Fig. \ref{fig:tunnelsketch}b). This process is then called spin-flip co-tunneling, and leads to a spin relaxation rate described by \cite{Qassemi2009, Lai2011, Otsuka2017}
\begin{equation}\label{eq:cot}
\Gamma_{\rm CT} = \frac{E_z}{\pi\hbar}\cdot\left(\frac{\hbar\Gamma_0}{e\alpha_{\rm p} V_{\rm p}}\right)^2. 
\end{equation}

Eq.~\eqref{eq:cot} shows that the co-tunneling rate is suppressed only quadratically (instead of exponentially) with plunge voltage, so it can in principle remain significant for $\mu_{\rm d,\uparrow} \lesssim \mu_{\rm SET}$. Eventually, when $\mu_{\rm d}\ll \mu_{\rm SET}$ all tunnel process should be suppressed (Fig. \ref{fig:tunnelsketch}c).

However, $\Gamma_{\rm CT}$ also depends quadratically (instead of linearly) on the bare tunnel rate $\Gamma_0$. Experiments showing co-tunneling effects have been ones where the electron under study was strongly tunnel-coupled to the charge reservoir, typically in a quantum transport setup \cite{Lai2011,Zumbuhl2004}, which requires $\Gamma_0 \gtrsim 1\times 10^{9}$~s$^{-1}$. Here we have instead $\Gamma_0 \lesssim 1\times 10^{3}$~s$^{-1}$, making the co-tunneling process extremely weak.

In Fig.~\ref{fig:plungedependence} we present the measurements of the spin relaxation rate $1/T_1$ as a function of plunge voltage $V_{\rm p}$. Fig.~\ref{fig:plungedependence}a shows the measured plunge voltage points with respect to $\mu_{\rm SET}$ (dashed purple line) in the charge stability diagram. We measure along two directions in the diagram: one with ``compensated'' plunging, i.e. moving $\mu_{\rm d}$ while keeping $\mu_{\rm SET}=const.$ using $\beta_{\rm c}=-0.51$ (green points, $V_{\rm p}^{\rm c}$), and one with ``uncompensated'' plunging perpendicular to the previous one using $\beta_{\rm uc}=1.82$ (blue points, $V_{\rm p}^{\rm uc}$). The latter allows for much higher $V_{\rm p}$ but also shifts $\mu_{\rm SET}$ and leads to a change in SET electron number $N$ when a Coulomb peak is crossed. 

As expected, the relaxation rate strongly decreases the deeper the donor is plunged below $\mu_{\rm SET}$ (Fig. \ref{fig:plungedependence}b), until it stabilises at around $T_1^{-1}=9^{-1}\,$s$^{-1}$. Clearly we identify two regimes: On the one hand, at high plunge voltages ($V_{\rm p}^{\rm c/uc}\gtrsim10\,$mV) the relaxation rate shows no dependence on $V_{\rm p}^{\rm c/uc}$, which means that the relaxation rate is not influenced by any type of tunneling process. On the other hand, at low $V_{\rm p}^{\rm c/uc}$, the relaxation strongly depends on $V_{\rm p}^{\rm c/uc}$. Fig.  \ref{fig:plungedependence}d shows the region $V_{\rm p}^{\rm c}\in(0,13)\,$mV in greater detail. 

To relate $V_{\rm p}^{\rm c}$ to energy, we determine $\alpha_{\rm p}$ by measuring the Zeeman energy through spin-dependent tunnelling (Fig. \ref{fig:plungedependence}c). Therefore, we tune the read level and measure at which voltages $\ket{\uparrow}$ and $\ket{\downarrow}$ tunnel out of the donor. We relate 
\begin{equation}
\Delta \mu_{\rm d}=\mu_{\rm d, \uparrow}-\mu_{\rm d, \downarrow}=E_{\rm Z}.
\end{equation}
To perform this measurement, we apply the following pulse sequence (inset Fig. \ref{fig:plungedependence}c). We load an electron with a random spin state, bias the donor at the read level with voltage $V_{\rm p}^{\rm c}$ and finally empty it. During the whole pulse sequence, the SET current is measured. Then we repeat the pulse sequence  while varying $V_{\rm p}^{\rm c}$ from $\mu_{\rm d, \downarrow}>\mu_{\rm SET}$, causing a high current by lifting Coulomb blockade regardless of the spin state, to $\mu_{\rm d, \uparrow}<\mu_{\rm SET}$, blocking conduction fully. In the intermediate regime where $\mu_{\rm d, \downarrow}\leq\mu_{\rm SET}\leq\mu_{\rm d, \uparrow}$, $\ket{\uparrow}$ tunnels to the SET, creating a current spike, and is replaced by $\ket{\downarrow}$ - we observe a spin tail \cite{Morello2010}. The voltage range of this tail $\Delta V_{\rm p}^{\rm c}=7.2\,$mV corresponds to the Zeeman energy at the external magnetic field of $B_0=5\,$T. From this we calculate the lever arm as 
\begin{equation}
\alpha_{\rm p}=\frac{E_{\rm Z}}{\Delta \mu_{\rm d}}=\frac{g\mu_B B_0}{e \Delta V_{\rm p}^{\rm c}}=0.08.
\end{equation} 
The voltage corresponding to the Zeeman energy at $1\,$T is thus $V_{\rm p}^{\rm Z}=1.4\,$mV, as indicated in Fig. \ref{fig:plungedependence}d. We indicated in the figure half the Zeeman energy, since this is the plunge voltage where $\mu_{\rm d, \uparrow}\geq\mu_{\rm SET}$. 

Within the detailed region in \ref{fig:plungedependence}d, we can again identify two regimes: For $V_{\rm p}\lesssim 2\,$mV, we observe a strong dependence of the relaxation rate on $V_{\rm p}^{\rm c}$, which we attribute to direct tunnelling from $\ket{\uparrow}$ to the SET reservoir. The purple line shows the predicted relaxation rate from Eq. \eqref{eq:directt} with use of realistic experimental parameters $\Gamma_0=50\,$~s$^{-1}$, $T_e=250\,$mK, and $V_{\rm p}^{\rm Z}=1.4\,$mV, $\alpha_{\rm p}=0.08$, as determined by the spin tail measurement.  

For $V_{\rm p}^{\rm c}\in(2,4)\,$mV we observe a slower decrease of the relaxation rate, which might indicate the  transition to a spin-flip co-tunneling mechanism. However, due to the slow direct tunneling rate $\Gamma_0=50\,$~s$^{-1}$, Eq. \eqref{eq:cot} predicts an extremely slow co-tunnelling rate $\Gamma_{\rm CT}(V_{\rm p}^{\rm c}=2\,\rm{mV})=3\times 10^{-9}\,$~s$^{-1}$. This rules out co-tunnelling for this relaxation process and leaves us searching for an explanation.

In order to fit the data with Eq. \eqref{eq:cot} we would have to assume $\Gamma_0=1.4\times 10^{6}\,$~s$^{-1}$ (dotted red line in Fig. \ref{fig:plungedependence}d), five orders of magnitude larger than the value extracted from the direct tunneling fit. This is could indicate that the region of slow decrease in $1/T_1$ has nothing to do with co-tunneling, or that the electron is able to virtually tunnel to some other charge center with a much larger bare tunnel rate, but the direct tunneling to this other center does not appear in the experiment withing the explored gate space. 

\begin{figure}
\centering
\includegraphics[width=\columnwidth]{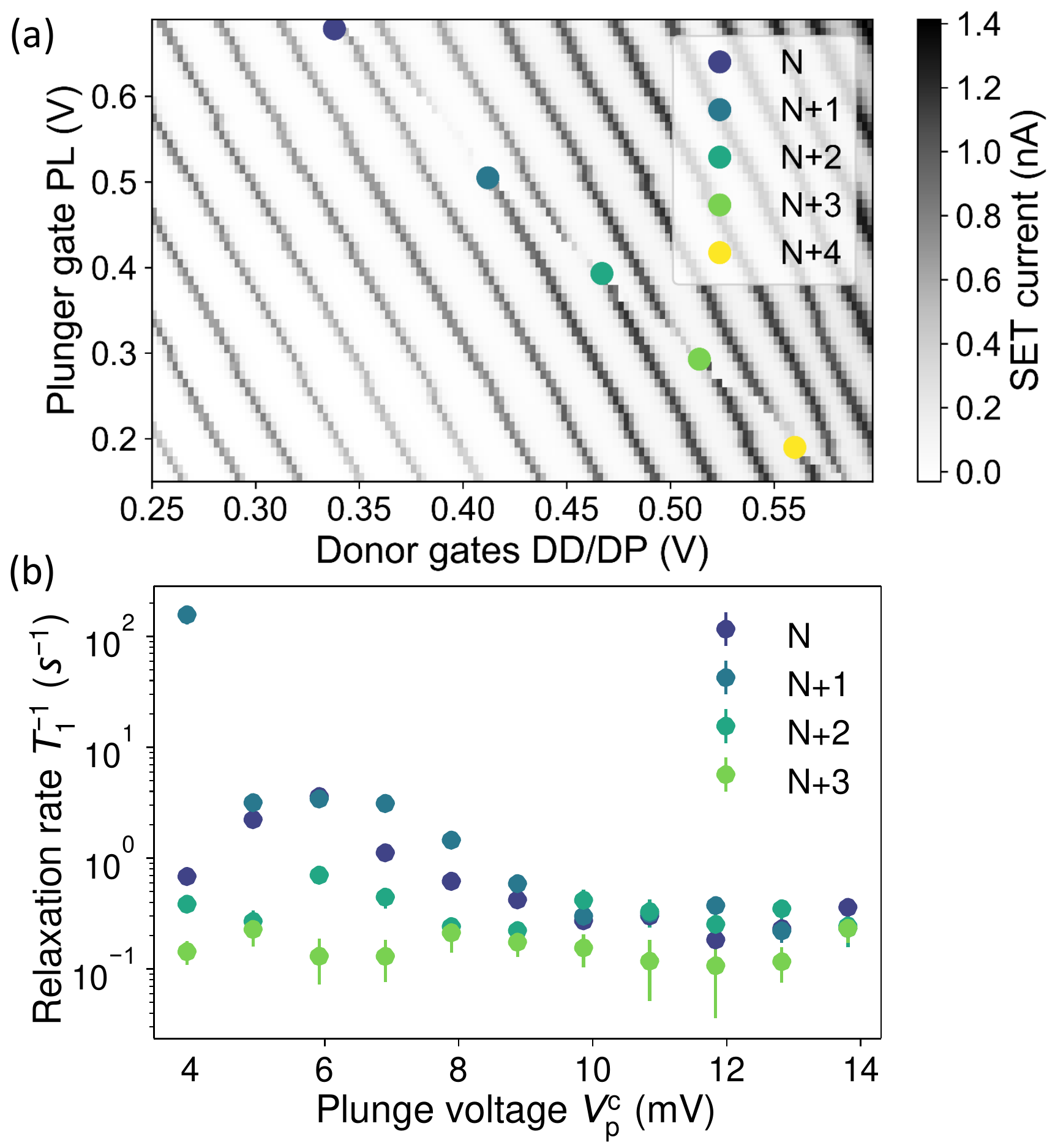}
\caption{(Color online) \textbf{Relaxation rate as a function of electron number. }
(a) Charge stability diagram as a function of plunger gate (PL) and donor gates (DD, DP) with bias points corresponding to different SET island electron numbers $N$ indicated. (b) Relaxation rates with plunge voltage for different $N$. Only compensated plunging is performed here. 
}
\label{fig:electronnumberdependence}
\end{figure}

We also study the relaxation time for several different SET Coulomb peaks, corresponding to a different electron number $N$ in the SET island. (Fig. \ref{fig:electronnumberdependence}). This experiment was performed after a thermal cycle of the device, resulting in a device tuning different from that in Fig.  \ref{fig:plungedependence}.  We find a strong variation in relaxation behaviour between different electron numbers for $V_{\rm p}^{\rm c}\lesssim 10\,$mV, when first-order tunneling processes are relevant. This is because the direct tunnel rate $\Gamma_{\rm DT}$ depends on the density of available states in the SET island. We estimate that our SET contains 100 electrons, which places it in an intermediate regime where it does not yet behave like a proper metallic electron reservoir with a continuous density of states, but shows some residual many-electron quantum-dot behaviour \cite{Nazarov2009}. As a consequence, the density of states is modulated by quantum effects arising from the Hund's rule when consecutively filling the electron orbitals. Different orbitals correspond to different wave functions, resulting in a different direct tunnel rate $\Gamma_{\rm DT}$ as a function of the SET electron number (at constant $\mu_{\rm d}$). In principle, the bare tunnel rate $\Gamma_0$ is modulated by density of state effects also as a function of $\mu_{\rm d}$, but the exponential influence of the Fermi function tends to mask this effect in the measurement of $1/T_1$. The non-uniform density of states in the SET island is most clearly visible in the spin tail measurement (Fig. \ref{fig:plungedependence}c), which clearly shows modulations of the a spin-up probability within the Zeeman energy window.

\section{\label{sec:conclusion}Conclusions}
In this work we have presented an extensive experimental study of the electron spin relaxation rate $1/T_1$ of single $^{31}$P donors, implanted in silicon metal-oxide-semiconductor devices. In particular, we have sought to highlight the subtle ways in which the presence of gating structures, metallic surfaces, crystal strain and tunnel coupling to charge reservoirs can make the relaxation rate deviate from that of bulk donors at equivalent temperatures and magnetic fields.

We found that Evanescent-Wave Johnson Noise (EWJN) is a likely candidate for the anomalous increase of the spin relaxation rate at low magnetic fields ($B \lesssim 1$~T) if the qubit is close to a highly conducting surface, such as the aluminum gates used here for electrostatic control of the donor. 

By analyzing $1/T_1$ of different devices in the regime where it is controlled by spin-phonon relaxation, we further deduced that lattice strain at the donor site might contribute to decreasing the relaxation rate, leading to very long $T_1$ times of up to $9.8$ seconds at $B=1\,$T. 

Finally, we analyzed the extent in which electron tunneling effects influence the spin relaxation, particularly when the donor electrochemical potential is in the vicinity of the Fermi level of an electron reservoir. The significance of this observation is that, when conducting experiment of very long duration (e.g. long dynamical decoupling sequences \cite{Muhonen2014}), it is essential to ensure that the donor potential is plunged well below the Fermi level. 

These observations will help designing and optimizing future devices to ensure that the spin relaxation time does not become a limit to the spin coherence time. We also hope they will stimulate further study of the microscopic origins of spin relaxation in realistic semiconductor devices, beyond the physics of bulk donors. 

\ \\

\begin{acknowledgements}
We thank W.A. Coish and V. Premakumar for helpful discussion, and J.J.L. Morton for providing the bulk data point in Fig.~\ref{fig:magnetic field dependence}. 

The research at UNSW and U. Melbourne was funded by the Australian Research Council Centre of Excellence for Quantum Computation and Communication Technology (Grants No. CE110001027 and No. CE170100012) and the US Army Research Office (Contracts No. W911NF-13-1-0024 and No. W911NF-17-1-0200). We acknowledge support from the Australian National Fabrication Facility (ANFF) and from the laboratory of Prof. R. Elliman at the Australian National University for the ion implantation facilities. KMI acknowledges support from Grant-in-Aid for Scientific Research by MEXT. The research of RJ was sponsored by the Army Research Office (ARO) under grant numbers W911NF-17-1-0274 and W911NF-12-1-0607. The views and conclusions contained in this document are those of the authors and should not be interpreted as representing the official policies, either expressed or implied, of the ARO or the US Government. The US Government is authorized to reproduce and distribute reprints for government purposes notwithstanding any copyright notation herein.
\end{acknowledgements}

%\bibliography{T1_bib} 


\end{document}